\documentclass[aps,pra,twocolumn,longbibliography]{revtex4-2}

\usepackage{amsmath}
\usepackage{amssymb}
\usepackage{graphicx}
\usepackage{booktabs}
\usepackage{hyperref}

\begin{document}
\title{Analytical model for structured light propagation\\ through a turbulent atmosphere}

\author{Konstantin Kravtsov}\email{konstantin.kravtsov@tii.ae}
\affiliation{
Technology Innovation Institute, Abu Dhabi, UAE
}

\date{\today}
\begin{abstract}
	We develop a straightforward analytical framework for the propagation of spatial
	light modes through a turbulent atmosphere. Built upon the split-step
	approach with the mode-based optical field representation, it
	directly assesses how turbulence-induced phase fluctuations
	deplete the optical power in the original mode and re-distribute it into neighboring spatial
	modes. Importantly, this power transfer is effectively a discrete diffusion
	process on a set of transverse modes, with
	the propagation distance serving as the evolution variable, yielding a simple
	solution for arbitrary distances in uniform channels in the
	form of a matrix exponential.  The transfer rate is determined by the spatial
	spectral overlap between the turbulence spectrum and the acceptance spectrum for a
	pair of interacting spatial modes. The model predicts the average power in each
	spatial mode and is exact when a single mode strongly dominates all others.
	Our predictions show reasonably good agreement with simulations up to
	medium-to-strong turbulence levels. The model also confirms the scalings with
	mode order previously known from empirical observations.
\end{abstract}
\maketitle

\section{Introduction}

Free-space optical communications are becoming increasingly important in today's economy.
Especially with the development of {\em quantum} communications, there is a growing
demand for the transmission of high-dimensional quantum states~\cite{MMS15,SBF17}.
The spatial degree of freedom, i.e.,  spatial {\em structuring} of light, is
a perfect resource for both high-dimensional state transmission~\cite{CLB19} and channel
multiplexing~\cite{M00}.

Nevertheless, the value of this resource may be limited by unavoidable imperfections in
the atmospheric channel, which arise from turbulent processes in the air.
They cause spatial mode crosstalk and loss of signal in a particular mode.

Communications with structured light in atmospheric channels were extensively studied in a
large number of publications, including those focusing on orbital angular momentum (OAM)
modes~\cite{P05, TB09, RLM12,BKI24}. The degradation of spatial quantum states in such channels
was also investigated both experimentally~\cite{HRM13} and in simulations~\cite{LSB15,AKF16}.
The performance of different types of spatial modes was analyzed in order to optimize the
properties of the channel~\cite{NMM17}. References to many more related studies can be
found in review papers~\cite{CMN21,PCF25}.

Despite being well developed, this field has historically been centered on empirical
observations and the extensive use of simulation tools, from early
examples~\cite{MF88,B97} to more recent ones~\cite{KS23,PCF25,BIS25}. To our knowledge, if we exclude
publications that rely on explicit random phase-screen generation, we are left with only a
modest number of works. First, it is the mode-coupled theory developed
in~\cite{ZZL15,ZZL16,ZWZ18}. It shares part of the underlying concept with the present
work, but was never extended to a complete set of modes. Thus, it could not provide a
comprehensive picture. Second, there is an analytical approach based on the quadratic
approximation to the Kolmogorov structure function~\cite{RWS15}. Unfortunately, as its
authors acknowledge, it could not be extended to other spectral profiles. Third, there is
an analytical paper~\cite{BSB19} that provides an explanation to the observed universal
scaling of the entanglement decay also studied in Section~\ref{sec_diagonal_scaling}.
Finally, there is our previous work~\cite{KZR18} that describes the performance of a single spatial mode in a turbulent
environment based on the lowest order term in a series expansion of a turbulent
perturbation.
It helped to obtain analytical expressions for channel properties.

The present work is an accurate analytical study of the same problem that leads to exact
results for a broad range of cases. In particular, it studies free-space optical channels
excited at the input with a single spatial mode. Propagation in the turbulent medium
causes modal crosstalk, so the optical power leaks from the original mode into other available
ones. The model provides the expected average power distribution among the modes.

In the present study, we focus only on {\em scalar} spatial modes. The same may be applied to
{\em vector} modes in a straightforward way, as the latter are superpositions of the
former in two non-interacting polarization domains~\cite{CRL16}.
Our results for Gaussian modes with two different types of symmetry
could also be adapted to other types of spatial modes. However, the general trends of mode
propagation in turbulent environments are the same for arbitrary mode types, cf.~\cite{MBF18}.

The paper is organized as follows. Section~\ref{seq_turbulence} reviews key results
on atmospheric turbulence and light propagation. In Section~\ref{seq_framework} we
define our framework and provide the main results. Section~\ref{seq_gaussian} specifically
discusses propagation of a Gaussian beam as a representative example. In
Section~\ref{sec_turb_applicability} we analyze the applicability of the model and its
specific details.

\section{Selected results on light propagation in turbulence}\label{seq_turbulence}
Turbulent distortions of light propagating through a turbulent medium originate from tiny
refraction index fluctuations around its mean value. The major result derived by
Kolmogorov and confirmed by many experimental studies is that the structure function of
the refraction index $n$ in a uniform and isotropic turbulent atmosphere (in 3 dimensions) scales with
the distance $r$ as
$r^{2/3}$~\cite{B93}:
\begin{equation}\label{eq_kolm_strf_n}
	D_n(r)\equiv\Bigl< {\bigl(n(r) - n(0)\bigr)}^2\Bigr> = C_n^2 r^{2/3},
\end{equation}
where $C_n^2$ is the so-called refractive index structure constant.
Switching to the Fourier domain, defined by the corresponding wavenumber $K$,
the previous expression is equivalent to
the spatial power spectrum $\Phi_n(K)$ in the form
\begin{equation}\label{eq_kolm_spec_n}
	\Phi_n(K)=\frac5{18\pi\,\Gamma(1/3)}C_n^2K^{-11/3}\approx 0.033\,C_n^2 K^{-11/3},
\end{equation}
where $\Gamma(x)$ is the gamma function.
Strictly speaking, the given scaling is observed only for some range of distances $r$
(and, correspondingly, wavenumbers $K$ in the Fourier domain) between the inner $(l_0)$ and
the outer $(L_0)$ scales of turbulence~\cite{LS70}, where $l_0 < L_0$.
However, we will ignore it for now.

The next important result is a transition from refractive index perturbations to
fluctuations of optical phase $S$ for a paraxial optical beam with wavelength $\lambda$
and wavenumber $k=2\pi/\lambda$ after the propagated distance of $L$. The derivation of
the optical phase structure function $D_S(\rho)$ and its power spectrum $F_S(K)$ in a
2-dimensional plane perpendicular to the light propagation direction yields~\cite{B93}
\begin{multline}\label{eq_strfun_integral}
	D_S(\rho)\equiv\Bigl< {\bigl(S(\rho) - S(0)\bigr)}^2\Bigr> =\\
	4\pi \int_0^\infty K\bigl[1-J_0(K\rho)\bigr]F_S(K)\,dK,
\end{multline}
where $J_0$ is the Bessel function of the first kind, and the 2-D transverse distance
$\rho$ is explicitly used instead of the 3-D distance $r$ to avoid ambiguity.
Notably, the 2-D phase
fluctuation spectrum coincides with the 3-D spectrum of the refractive index fluctuations
up to a constant:
\begin{equation}\label{eq_gen_phase_spectrum}
	F_S(K) = 2\pi k^2 L \Phi_n(K) \approx 0.207 C_n^2 k^2 L K^{-11/3}
\end{equation}
The evaluation of the integral~(\ref{eq_strfun_integral}) yields
\begin{equation}
	D_S(\rho) = 2\frac{\sqrt{\pi}\:\Gamma(1/6)}{5\Gamma(2/3)}C_n^2 k^2 L
\rho^{5/3}\approx
	2\times 1.46 C_n^2 k^2 L \rho^{5/3}.
\end{equation}

Another conventional form of the expression for the phase structure function was
proposed by Fried~\cite{F66}
\begin{equation}
	D_S(\rho) = 8\sqrt{2}\left[\frac 35 \Gamma\left(\frac 65\right)\right]^{5/6}
	\left(\frac{\rho}{r_0}\right)^{5/3} \approx 6.88
\left(\frac{\rho}{r_0}\right)^{5/3},
\end{equation}
where $r_0$ is the Fried parameter. In a sense, $r_0$ defines the best angular resolution
achievable when observing through a turbulent channel: it corresponds to the
diffraction-limited resolution of a telescope with an aperture diameter equal to
$r_0$~\cite{R81}.
Comparing the last two expressions gives a definition of the Fried parameter
\begin{equation}\label{eq_fried_definition}
	r_0 = \left(0.42C_n^2k^2 L\right)^{-3/5}.
\end{equation}

Looking at given expressions for structure functions (or spectra), we can see that they
diverge as $\rho\rightarrow\infty$
(or $K\rightarrow0$). Formally, this means that at an arbitrary point in the transverse
plane, the phase variance
given by the integral $(2\pi)^{-2}\iint_0^\infty F_S(K)\,dK_x\,dK_y$ is infinite, since
the integral
diverges as $K\rightarrow0$. In fact, this is the reason why a more complex description
based on the structure function is chosen instead of a more conventional covariance-based
approach. This divergence is unphysical and stems from the continuation
of~(\ref{eq_kolm_strf_n})
beyond the applicable range of arguments from $l_0$ to $L_0$.

The problem is resolved by attenuating the spectrum of fluctuations outside this
range. One common way to do this is to express the spectrum $\Phi_n$ or $F_S$ via the
spatial frequency
$f=K/(2\pi)$ and substitute
\begin{equation}\label{eq_mod_vonkarman_def}
	f^{-11/3} \longrightarrow \frac{\exp(-l_0^2 f^2)}{\left(f^2+
	L_0^{-2}\right)^{11/6}},
\end{equation}
which corresponds to the so-called modified von Karman turbulence model~\cite{B93}.

Finally, another important quantity that will appear in the present paper defines the
strength of turbulence for a given optical channel and is called the Rytov parameter
$\sigma_R^2$.
It is defined as the variance of $\ln(I/I_0)$ for low turbulence levels,
where $I$ is the intensity observed
at a given point in the output plane of the channel, assuming that the input plane
contains an ideal
plane wave; and $I_0$ is an arbitrary reference intensity. As derived in~\cite{T71} for
the spectrum~(\ref{eq_kolm_spec_n})
\begin{equation}\label{eq_rytov_definition}
	\sigma_R^2 = 4\,\frac{2^{1/6}\pi^{3/2}(\sqrt{3}-1)}{11\,\Gamma(2/3)} C_n^2
	k^{7/6}L^{11/6} \approx 1.23 C_n^2 k^{7/6}L^{11/6}.
\end{equation}
For larger $\sigma_R^2 \gtrsim 0.5$ its connection with log-intensity fluctuations breaks,
as the experimentally measured fluctuations tend to saturate around $\sigma_{ln\,I}^2
\approx 1$~\cite{LS70,B93}.

\section{Modal analysis framework}\label{seq_framework}
\begin{figure}[tb]
\begin{center}
\includegraphics[width=\columnwidth]{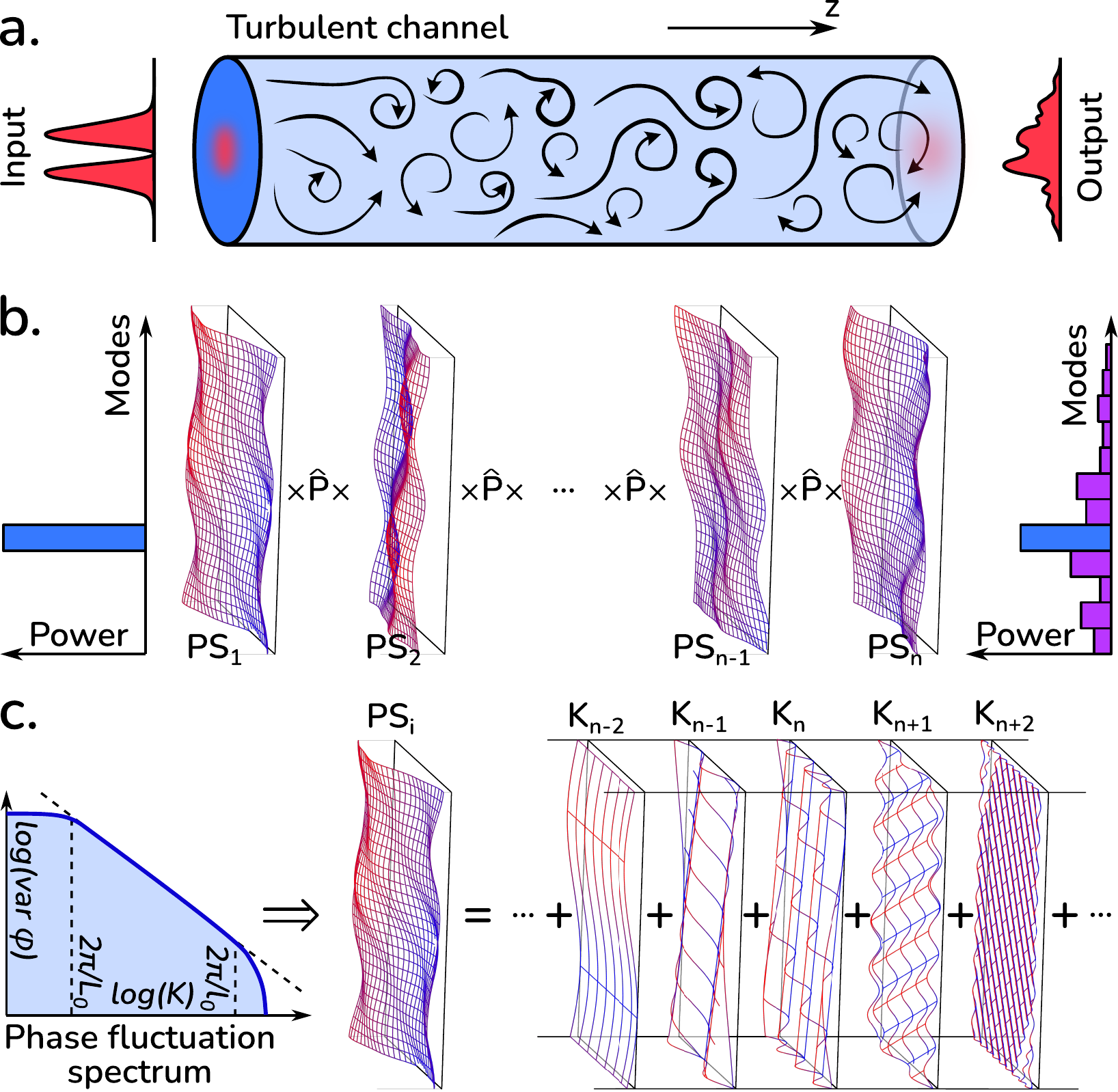}
\end{center}
\caption{\label{fig_t_concept}
	Key elements of the proposed framework.\\{\bf a.} Schematic of the implied physical
	setup: a single spatial mode of light enters an extended channel with a turbulent
	medium, where it gets distorted. {\bf b.} The same problem in the split-step
	representation: the channel is divided into a large number of random and independent
	chunks. Each chunk is modeled as a zero-thickness phase screen (PS$_i$) and a free
	propagation region defined by the propagator $\hat{P}$. The resulting field in the
	end of the channel is decomposed into spatial modes to analyze their statistics.
	Our key proposal is to use the depicted {\em optical-power-versus-spatial-mode\/}
	encoding, which turns $\hat{P}$ into the identity operator;
	therefore, the overall effect of the channel is just a collective action of
	infinitesimal phase screens.
	{\bf c.} Each phase screen is defined by a given spatial spectrum of phase
	fluctuations, shown schematically on a base-10 log–log scale (left). It allows
	each infinitesimal phase screen to be represented as a sum
	of a large number of elementary phase waves with random directions and varying
	wavevectors $K_i$.
}
\end{figure}

We study the propagation of spatial modes of light in a turbulent medium, as
schematically depicted in Fig.~\ref{fig_t_concept}a. Following the conventional split-step
representation of the channel, we can divide it into a large number of infinitesimal
pieces as shown in Fig.~\ref{fig_t_concept}b. Each of them can be treated as a localized
phase perturbation and a corresponding free propagation region. The key element of the
proposed approach is focusing on discrete spatial optical modes rather than on the
entire electromagnetic field distribution. The modes themselves are eigenfunction
solutions of the propagation equation and allow one to avoid dealing with the propagation
equation of the electromagnetic field per se.
In particular, we only account for the optical power in each
of the modes as shown graphically at the input and output of the channel.

An apparent advantage of such an encoding is its invariance under free propagation.
Thus, one of the two components in the split-step model disappears, while only the accumulated
effects of infinitesimal phase screens on the modal distribution remain. This greatly
simplifies the problem and makes it analytically integrable, as we show below. At the same
time, this encoding ignores the relative phases between spatial modes, which has consequences
for the applicability of the model discussed in depth in
Section~\ref{sec_turb_applicability}.

Each infinitesimal phase screen follows the same shape of the spatial power spectrum (see
Fig.~\ref{fig_t_concept}c.) and can be perceived as a large set of independent phase
perturbations of different spatial frequencies. Therefore, we model the resulting effect
of a phase screen as a direct sum of elementary
harmonic phase distortions (waves) with certain spatial frequencies and amplitudes. Their
directions in the plane perpendicular to the light propagation direction and absolute
phases are random. Each individual perturbation is weak and independent of the others.

The induced phase fluctuations associated with each infinitesimal phase screen
break the orthogonality between spatial modes and lead to
power redistribution among the modes. First, we study this effect for an elementary
single-frequency phase wave contributing to an extremely weak phase screen.
Second, we combine contributions for the entire ensemble of those elementary waves with varying spatial
frequency to arrive at the contribution of an individual phase screen. Finally, an extended
channel is constructed from the infinite number of individual phase screen contributions.

We adopt the following notation: $z$ is the direction of the paraxial beam
propagation; $x$ and $y$ define the
perpendicular plane. All modes are centered on $x=y=0$. We also use the cylindrical
coordinate system $\rho$, $\theta$, $z$.

We explicitly study two convenient mode sets: Hermite-Gaussian (HG) and Laguerre-Gaussian
(LG), however any other orthogonal mode set can be used similarly. $G(x,y)$ (or
$G(\rho,\theta)$) is the mode field profile in the transverse plane; the indices $a$ and
$b$ fully specify the corresponding modes.

The orthogonality of the mode set indicates that
\begin{equation}
	\iint_{-\infty}^\infty G_a(x,y) G_b^*(x,y)\,dx\,dy = \delta_{ab},
\end{equation}
i.e., zero if $a\ne b$; where $*$ denotes the complex conjugation. The onset of a phase wave
with the wavenumber $K = \sqrt{K_x^2 + K_y^2}$ and the amplitude $\alpha_K$
\begin{equation}\label{eq_elem_phase_wave}
	\varphi_K(x,y) = \alpha_K\cos(K_x x + K_y y + \phi)
\end{equation}
breaks this orthogonality. So, the fraction of power that migrates from mode $a$ to mode
$b$ due to the perturbation (cf.\ the transition probability) is the squared modulus of
the
transition amplitude
\begin{multline}\label{eq_transition_prob_gen}
	P_{ab}(K) = |A_{ab}(K)|^2 = \\
	\left|\iint_{-\infty}^{\infty} G_a(x,y) G_b^*(x,y) \exp\bigl[i
	\varphi_K(x,y)\bigr]\,
	dx\,dy\right|^2.
\end{multline}
One can see that $P_{ab}(K)$ is independent of the mode ordering. In addition, if the particular
mode set is complete, power conservation requires that $\sum_b P_{ab}(K) = 1$.

Since each elementary single-frequency phase wave is a tiny contribution to the infinitesimal phase screen,
we use the following power series decomposition ignoring higher-order terms
\begin{equation}\label{eq_exponent_expansion}
	\exp\bigl[i \varphi(x,y)\bigr] = 1 + i \varphi(x,y) - \varphi^2(x,y) / 2 +
O(\varphi^3).
\end{equation}

Hermite-Gaussian modes may be described by the following expression that ignores the
global phase that is independent of the transverse coordinates
\begin{multline}\label{eq_hgmodes_definition}
	HG_{mn}(x,y) = \sqrt{\frac{k
z_R}\pi}\;\frac{\exp{\Bigl[-k(x^2+y^2)/2(z_R-iz)}\Bigr]}{\sqrt{2^{m+n}\,n!\,m!}\;
(z_R-iz)}\\
	\times H_m\left(x\sqrt{\frac{kz_R}{z_R^2 + z^2}}\right)
	H_n\left(y\sqrt{\frac{kz_R}{z_R^2 + z^2}}\right),
\end{multline}
where $z_R=\pi w_0^2 /\lambda$ is the Rayleigh length, $H_n$ is the Hermitian polynomial
of order $n$, $i^2 = -1$, and $z = 0$ defines the position of the beam waist.
Each mode is defined by a pair of non-negative integer indices $m$ and $n$, while their
sum is designated as $N$ and defines the mode order $N=m+n$.

Laguerre-Gaussian modes have cylindrical symmetry and thus are defined in the cylindrical
coordinates instead. Their transverse profile is given by
\begin{multline}
	LG_{pl}(\rho,\theta) = \sqrt{\frac{p!}{\pi (p+|l|)!}}
	\left(\frac{\sqrt{k z_R}}{z_R-iz}\right)^{|l|+1} \rho^{|l|} e^{i l\theta}\\
	\times L_p^{|l|} \left(\frac{k z_R \rho^2}{z_R^2+z^2}\right)
	\exp\left[-\frac{k \rho^2}{2(z_R-iz)}\right],
\end{multline}
where $L_q^s$ is the generalized Laguerre polynomial. The mode is defined by a
non-negative
integer $p$ and an integer $l$. The mode order for LG modes is $N= 2p+|l|$.

Each mode in one of the sets is a linear combination of modes of the same order $N$ in
the other set~\cite{DA92,RGG16}. Both mode sets have a distance-varying spot radius
\begin{equation}
	w(z) = w_0\sqrt{1+ (z/z_R)^2}.
\end{equation}
So, the parameter $w_0$ has the meaning of the characteristic waist radius, while $z_R$ is
one half of the waist length. In order to provide a dimensionless representation of the
involved calculus, we will extensively use the following dimensionless variable
\begin{equation}
	\vartheta_K = K^2\;\frac{z_R^2 + z^2}{4k z_R} = \frac{K^2 w^2(z)}8,
\end{equation}
which is roughly the number of full phase waves per mode diameter squared. To simplify the
notation, we will often omit the subscript for $\vartheta$ and the argument for $w$ below.

\subsection{Interaction of Laguerre-Gaussian modes}
The analog of~(\ref{eq_transition_prob_gen}) for LG modes in cylindrical coordinates is
\begin{multline}\label{eq_amp_lg_gen}
	A_{pl;qs}^{LG} = \\
	\int_0^\infty \rho\,d\rho \int_0^{2\pi} d\theta\;
	LG_{pl}(\rho,\theta) LG_{qs}^*(\rho,\theta) \exp\bigl[i\varphi(\rho,\theta)\bigr].
\end{multline}
Due to cylindrical symmetry, this quantity does not depend on the direction $(K_x,K_y)$
of the phase wave, so without loss of generality we may assume $K_x=K$, $K_y=0$, and the
argument of the $\cos$ function in~(\ref{eq_elem_phase_wave}) becomes $K\rho\cos\theta +
\phi$.

We evaluate integrals in~(\ref{eq_amp_lg_gen}) using (cf.~\cite[8.411.2-3]{GRXX})
\begin{multline}
	\int_0^{2\pi} e^{i(l-s)\theta} \cos(\kappa r \cos\theta+\phi)\,d\theta\\=
	\pm 2\pi J_{|l-s|} (\kappa r)
	\begin{cases}
		\cos\phi &\text{if}\; l-s \;\text{is even},\\
		\sin\phi &\text{otherwise},
	\end{cases}
\end{multline}
where $J_n$ is the Bessel function of the $n$-th order.
Then applying~\cite[7.422.2, to be corrected in future editions]{GRXX}
\begin{multline}
	\int_0^{\infty} x^{\nu+\sigma+1} e^{-x^2} L_m^\nu (x^2) L_n^\sigma (x^2)
	J_{\nu+\sigma} (2yx)\, dx \\
	= \frac{(-1)^{m+n}}{2} y^{\nu+\sigma} e^{-y^2} L_n^{m-n+\nu}(y^2)
	L_m^{n-m+\sigma}(y^2),
\end{multline}
and~\cite[7.422.3, to be included in future editions]{GRXX}
\begin{multline}
	\int_0^{\infty} x^{\nu+\sigma+1} e^{-x^2} L_m^\nu (x^2) L_n^\sigma (x^2)
	J_{\nu-\sigma} (2yx)\, dx \\
	= (-1)^{m+n} \frac{(n+\sigma)!}{2n!} y^{\nu-\sigma} e^{-y^2}
	L_{n+\sigma}^{m-n+\nu-\sigma}(y^2) L_m^{n-m}(y^2)
\end{multline}
we arrive at a closed form expression. Finally, when calculating $P_{pl;qs}(K)$ we average
the result over all possible phases $\phi$ of the phase wave, so $\langle
\cos^2\phi\rangle = \langle \sin^2\phi\rangle = 1/2$.

The evaluation of such overlaps between different spatial modes ($p\ne q$ or $l\ne s$) is
the most straightforward, since only the linear term in~(\ref{eq_exponent_expansion})
provides the dominant contribution. So, the resulting coupling is
\begin{multline}
	P_{pl;qs}^{LG}(K) =
	\frac{\alpha_K^2}{2}\frac{q!}{(q+|s|)!}
	{\vartheta_K}^{|s|} \exp(-2\vartheta_K)\\
	\times\begin{cases}
		\frac{(p+|l|)!}{p!} {\vartheta_K}^{-|l|}
		\left[
			L_{p+|l|}^{q-p+|s|-|l|}(\vartheta_K)
			L_q^{p-q}(\vartheta_K)
			\right]^2 &\text{if}\; ls\ge 0,\\
		\frac{p!}{(p+|l|)!} {\vartheta_K}^{|l|}
		\left[
			L_p^{q-p+|s|}(\vartheta_K)
			L_q^{p-q+|l|}(\vartheta_K)
			\right]^2 &\text{if}\; ls\le 0.
	\end{cases}
\end{multline}
Both expressions actually coincide if $ls = 0$, which is a nontrivial fact in the case
$l\ne 0$.

In the case of the same mode we have to deal with both the constant and the linear term,
with the resulting expression
\begin{multline}
	P_{pl;pl}^{LG}(K) = \\1 - \frac{\alpha_K^2}{2}\left(
	1 - \left[L_{p+|l|}(\vartheta_K)L_p(\vartheta_K)\right]^2
	\exp(-2\vartheta_K)\right)
\end{multline}

\subsection{Interaction of Hermite-Gaussian modes}

The evaluation of~(\ref{eq_transition_prob_gen}) for the HG modes proceeds in a similar
way. However, owing to the lack of rotation symmetry, the result depends on the direction of
the phase wave. In the following, we use this form: $K_x = K\cos\xi$ and $K_y = K\sin\xi$.

An essential evaluation~\cite[7.374.7]{GRXX}
\begin{multline}
	\int_{-\infty}^{\infty} e^{-(x-y)^2} H_m(x)H_n(x)\,dx\\
	=2^n \sqrt{\pi} m! y^{n-m}L_m^{n-m}(-2y^2),
\end{multline}
which is valid for any non-negative integers $m \le n$, is used for both $x$ and $y$
directions. As the final result is symmetric with respect to the exchange
$m \leftrightarrow k$ or $n \leftrightarrow t$, below we use the assumption that
$m \le k$ and $n \le t$ for certainty. We also perform averaging over phase $\phi$ in the
same way as for LG modes.

The result for a pair of distinct modes is as follows
\begin{multline}
	P_{mn;kt}^{HG}(K) =
	\frac{\alpha_K^2}{2}\frac{m!\,n!}{k!\,t!}
	{\bigl(2\vartheta_K\cos^2\xi\bigr)}^{k-m}
	{\bigl(2\vartheta_K\sin^2\xi\bigr)}^{t-n}\\
	\times{\Bigl[
		L_m^{k-m}\bigl(2\vartheta_K\cos^2\xi\bigr)
		L_n^{t-n}\bigl(2\vartheta_K\sin^2\xi\bigr)
		\Bigr]}^2 \exp(-2\vartheta_K).
\end{multline}

In the case of the same mode, the remaining power in the particular Hermite-Gaussian
mode is given by
\begin{multline}
	P_{mn;mn}^{HG}(K) = 1 - \frac{\alpha_K^2}{2}\times \\
	\left[ 1 - {\Bigl(
	L_m\bigl(2\vartheta_K\cos^2\xi\bigr)
	L_n\bigl(2\vartheta_K\sin^2\xi\bigr)\Bigr)}^2 \exp( -2\vartheta_K)\right].
\end{multline}

Finding these quantities for the isotropic case of real turbulence requires averaging
over all possible angles $\xi \in [0, 2\pi)$. At the time of writing, an analytical
approach
proved infeasible (see also Section~\ref{sec_diagonal_scaling} for analytical scalings).
For practical applications, numerical averaging should be used instead.

\subsection{Contribution of a single phase screen}

So far we have calculated the interaction of spatial modes induced by an elementary phase
wave of a certain wavenumber $K$ and the infinitesimal amplitude $\alpha_K$. The next step is to combine
such elementary perturbations with different wavenumbers as in the actual phase screen
produced by a thin layer of the turbulent atmosphere. The key condition is that, in the
actual turbulent process, contributions at different wavenumbers are mutually independent.
Therefore, the phase and direction of each elementary contribution are totally random,
matching the assumptions we made in our calculations earlier.

By performing direct integration, we have shown that the first
nonvanishing term of perturbation in the mode-coupling expressions is quadratic in
the amplitude of an elementary phase wave, regardless of the chosen mode family. Specifically,
the integration results have the form of
\begin{equation}
	P_{ab}(K) = \begin{cases}\alpha_K^2/2\; B_{ab}(\vartheta_K), &\mathrm{if}\; a\ne b,\\
	1 + \alpha_K^2/2\; B_{ab}(\vartheta_K), & \mathrm{if} \; a=b,
	\end{cases}
\end{equation}
where $B(\vartheta_K)$ is given by different expressions for the two studied sets of modes.
The diagonal elements $B_{aa}$ are negative, as each elementary wave depletes the power in
mode $a$, while the off-diagonal ones are positive, as a tiny part of the optical power in mode
$a$ is transferred to mode $b$.
The key insight is that the pre-factor in each of the expressions for the elementary
perturbation, $\alpha_K^2/2$, is, in fact, the variance of the phase deviation for an
elementary wave~(\ref{eq_elem_phase_wave}). However, phase variance is the additive
quantity that defines the power spectrum. Indeed, for two distinct $K$ and $K'$ due to
their independence, $\langle(\varphi_K + \varphi_{K'})^2\rangle = \langle \varphi_K^2\rangle +
\langle\varphi_{K'}^2\rangle$, so the contributions for different wavevectors are
additive.

The combined average effect produced by an ensemble of elementary phase waves with varying
$K$ is therefore obtained as the sum of their individual contributions. At the same time,
the phase variance associated with a small spectral width $\Delta K$ around the value
$K$ is defined by the 2-dimensional power spectrum $F_S(K)$ as $F_S(K)\,2\pi K\Delta K$.
Thus, the overall combined perturbation caused by a weak phase screen is given by
\begin{equation}\label{eq_pab_action_strength}
\int_0^\infty F_S(K) B_{ab}(\vartheta_K)\,2\pi K \,dK.
\end{equation}

Another insight is that $|B(\vartheta_K)| \le 1$, so it effectively defines a
mode-specific spectral filter that partially accepts the phase perturbation power into the coupling strength
between modes. Therefore, in what follows, we refer to $B_{ab}(\vartheta_K)$ as the
acceptance spectrum. The overall mode coupling produced by a weak phase screen is equal
to the total power of its phase fluctuations overlapping with the acceptance spectrum.

\subsection{General extended turbulent channel}
In an extended channel, the power of phase perturbations is also an additive quantity:
each chunk of the optical path adds a certain power to phase perturbations. This follows
from the linear dependence of the power spectrum $F_S(K)$ on the channel length $L$
in~(\ref{eq_gen_phase_spectrum}).
Therefore, the coupling rate is defined by the accumulation rate of phase fluctuations,
i.e.,\ the derivative $dF_S(K,L)/dL$. Another simplification is to use dimensionless $\vartheta_K$ instead of $K$ as the
argument of the phase perturbation spectrum. Combining these two ideas, we define the
following quantity:
\begin{multline}\label{eq_lambda_definition}
\Lambda_{ab} \equiv \int_0^\infty \frac{dF_S(K,L)}{dL} B_{ab}(\vartheta_K)\, 2\pi K\,dK\\
=\frac{8\pi}{w^2}\int_0^\infty \frac{dF_S(\vartheta,L)}{dL} B_{ab}(\vartheta)\, d
\vartheta.
\end{multline}

Thus, a weak phase screen associated with a small chunk of the channel having
length $\Delta L$ produces the following mode coupling:
\begin{equation}
	P_{ab} = \begin{cases}\Lambda_{ab}\,\Delta L, &\mathrm{if}\; a\ne b,\\
	1 + \Lambda_{ab}\,\Delta L, & \mathrm{if} \; a=b,
	\end{cases}
\end{equation}
Apparently, $\Lambda_{ab}$ define a matrix, which acts on the mode distribution as
$d\mathbf{v}/dL = \mathbf{\Lambda} \mathbf{v}$,
where $\mathbf{v}$ is the vector of optical powers in all
considered modes.

Similar equations appear in many disciplines and, depending on the
context, may be referred to as master equations, diffusion on a graph, continuous-time
Markov chains, discrete state kinetics, or kinetic schemes~\cite{J97,SWM15,OH23}. As the net exchange flux between any two modes is
proportional to the difference in their populations, the entire stochastic evolution is effectively a relaxation
process that tends to equalize the modal populations. Given that the total power is conserved,
it is natural to describe
this process as {\em diffusion\/} in the space of transverse modes, or as {\em
discrete diffusion\/} on a set of spatial modes. In the single-photon
case, the expression becomes the master equation governing the probabilities of finding
the photon in a given mode.

The accurate solution of this discrete diffusion problem for a uniform channel with an arbitrary length $L$ is given by the matrix exponent
\begin{equation}\label{eq_matrix_exponent_turb}
\mathbf{v}=\exp(\mathbf{\Lambda} L)\,\mathbf{v}_0,
\end{equation}
which is the final result of the present study. This expression has its own limits of
applicability, which are discussed in Section~\ref{sec_turb_applicability}.
For a non-uniform channel in which either the turbulence strength or the mode width $w$
varies with $z$, the diffusion/master equation must be integrated using a position-dependent
matrix $\mathbf{\Lambda}(z)$.

\subsection{Explicit results for the modified von Karman turbulence
spectrum~(\ref{eq_mod_vonkarman_def})}
Using $K^{-11/3}=w^{11/3}\vartheta^{-11/6}/(32\sqrt{2})$ and
substituting~(\ref{eq_gen_phase_spectrum}) into~(\ref{eq_lambda_definition}) with the von
Karman spectral damping, one can show
that in a uniform atmospheric channel
\begin{multline}\label{eq_full_vk_integral}
\Lambda_{ab} = 0.115 C_n^2 k^2 w^{5/3}\\
	\times\int_0^\infty {\left(\vartheta+\frac{\pi^2 w^2}{2 L_0^2}\right)}^{-11/6}\\
\times\exp\Bigl(-2\vartheta\, l_0^2 /(\pi^2w^2)\Bigr) \,B_{ab}(\vartheta)\,d\vartheta.
\end{multline}
Alternatively, expressing the strength of the turbulence in the form of the Fried
parameter, one can find that
\begin{equation}\label{eq_fried_for_strength}
	\Lambda_{ab}\,L = 0.272\left(\frac{w}{r_0}\right)^{5/3} \mathcal{I}_{ab},
\end{equation}
where $\mathcal{I}_{ab}$ is the same dimensionless integral over the entire spectrum that appears
in~(\ref{eq_full_vk_integral}).

\section{Power depletion in the fundamental mode}\label{seq_gaussian} 
An important practical topic is the propagation of a Gaussian beam through the turbulent
atmosphere. It can be analyzed in detail within the developed framework by looking at
interactions of the 00 mode with higher order ones. As discussed
earlier, each chunk of the propagation distance contributes a share of phase fluctuation
power that leads to re-distribution of a fraction of the optical power from the
fundamental mode into higher order modes.

For LG modes using $\bigl[L_q^{-q}(\vartheta)\bigr]^2 = \vartheta^{2q}/(q!)^2$ it is
convenient to re-write the acceptance spectrum for a group of modes with mode order $N$ as
\begin{equation}
B_{00;\min(j,N-j), N-2j}^{LG}(\vartheta) = \frac{\vartheta^{N}\exp{(-2\vartheta)}}
{j!(N-j)!},
\end{equation}
where integer $j\in \{0,1,\dots,N\}$ indexes all modes in the group. Thus, the
generation rates for LG modes of a given order $N$ are proportional to the $N$th row of Pascal’s
triangle:
the maximal rates are observed for central modes with minimal (or zero) OAM; in contrast,
the smallest rates correspond to the modes with the highest azimuthal index.

For the HG family, we also consider a group of modes of order $N=m+n$. First, we can
perform directional averaging using
\begin{equation}
	\left< \cos^{2m}\xi \sin^{2n}\xi\right> =
\frac{(2m)!\,(2n)!}{4^{m+n}\,(m+n)!\,m!\,n!}.
\end{equation}
Then, the resulting acceptance spectrum becomes
\begin{equation}
	B_{00;mn}^{HG}(\vartheta) =
	\frac{\vartheta^N\exp{(-2\vartheta)}}{2^N\,N!}
	{2m \choose m}{2n \choose n}.
\end{equation}
This distribution within the group is much more uniform than for LG modes, with some advantage for the
generation of highly asymmetric modes where $m$ or $n$ approach $N$.

The acceptance spectrum for the power remaining in the fundamental mode (which is the same
in both mode families) is negative and equals
\begin{equation}\label{eq_bzerozero}
B_{00,00}(\vartheta) = \exp(-2\vartheta) - 1.
\end{equation}

It is straightforward to show that the combined power loss towards the group of modes of
a certain order $N>0$ is
\begin{equation}\label{eq_bzerononzero}
B_{00,N}(\vartheta) = \frac{(2\vartheta)^N}{N!}\exp(-2\vartheta)
\end{equation}
regardless of the chosen mode type. This result for HG modes may be obtained using
\begin{equation}
	\sum _{k=0}^{n}{\binom {2k}{k}}{\binom {2n-2k}{n-k}}=4^{n}.
\end{equation}
As follows from~(\ref{eq_bzerozero}) and~(\ref{eq_bzerononzero}), the total power is always conserved as $\sum_{N=0}^\infty B_{00,N} = 0$.

Substituting all acceptance spectra into~(\ref{eq_full_vk_integral}) one can find the
entire matrix of mode connections $\mathbf{\Lambda}$ that defines the average dynamics of the system.
\begin{figure}[!htbp]
\begin{center}
\includegraphics[width=0.9\columnwidth]{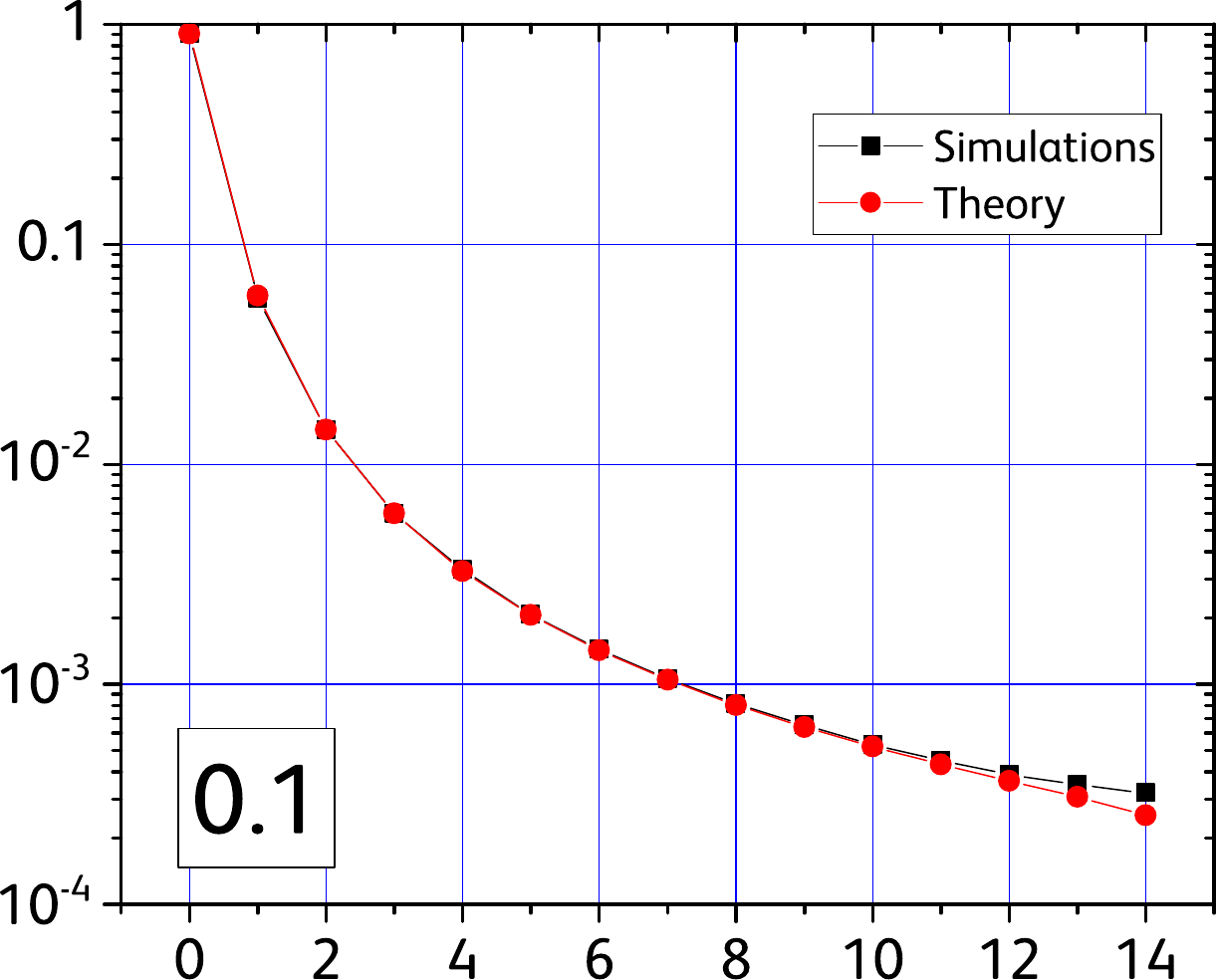}
\includegraphics[width=0.9\columnwidth]{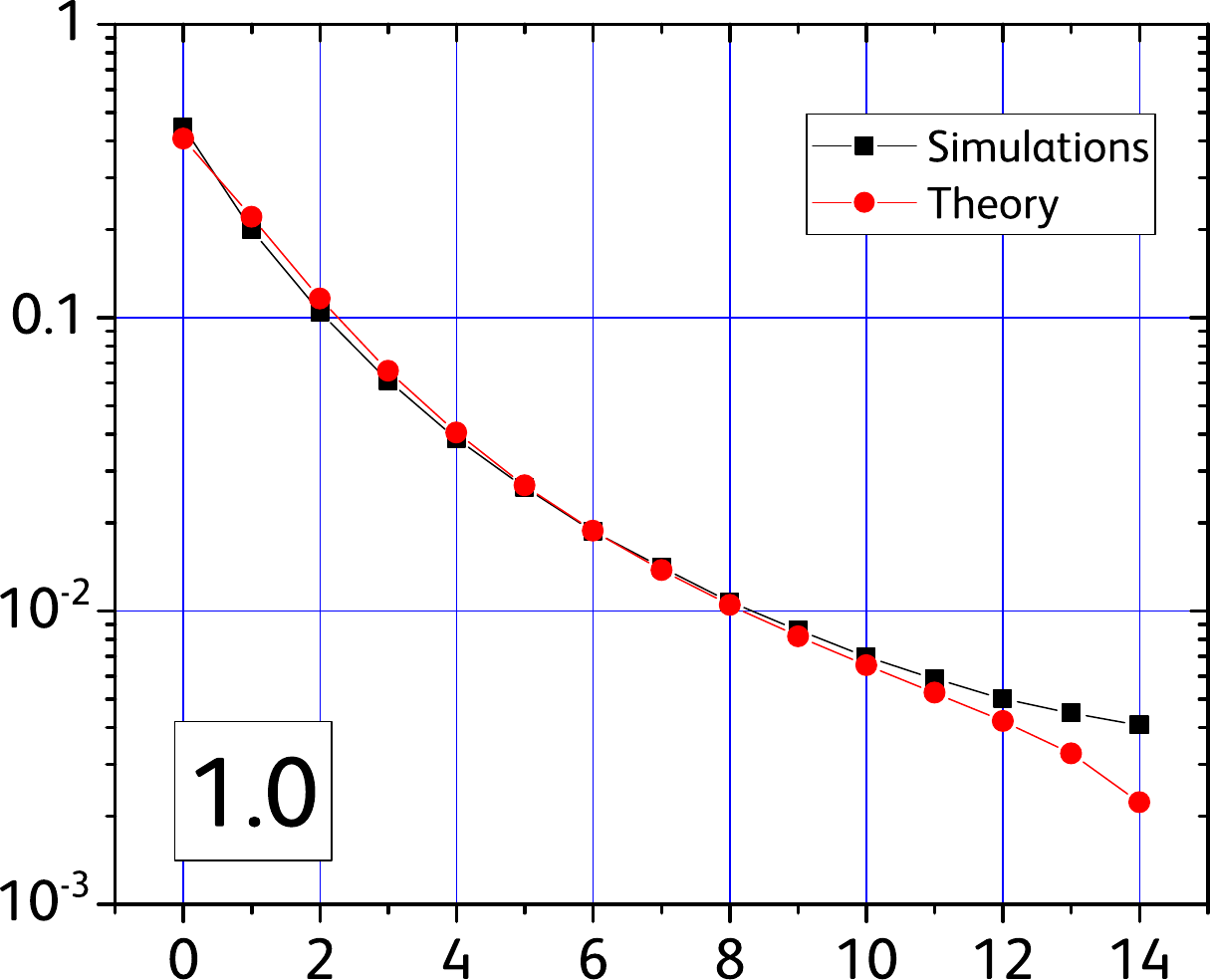}
\includegraphics[width=0.9\columnwidth]{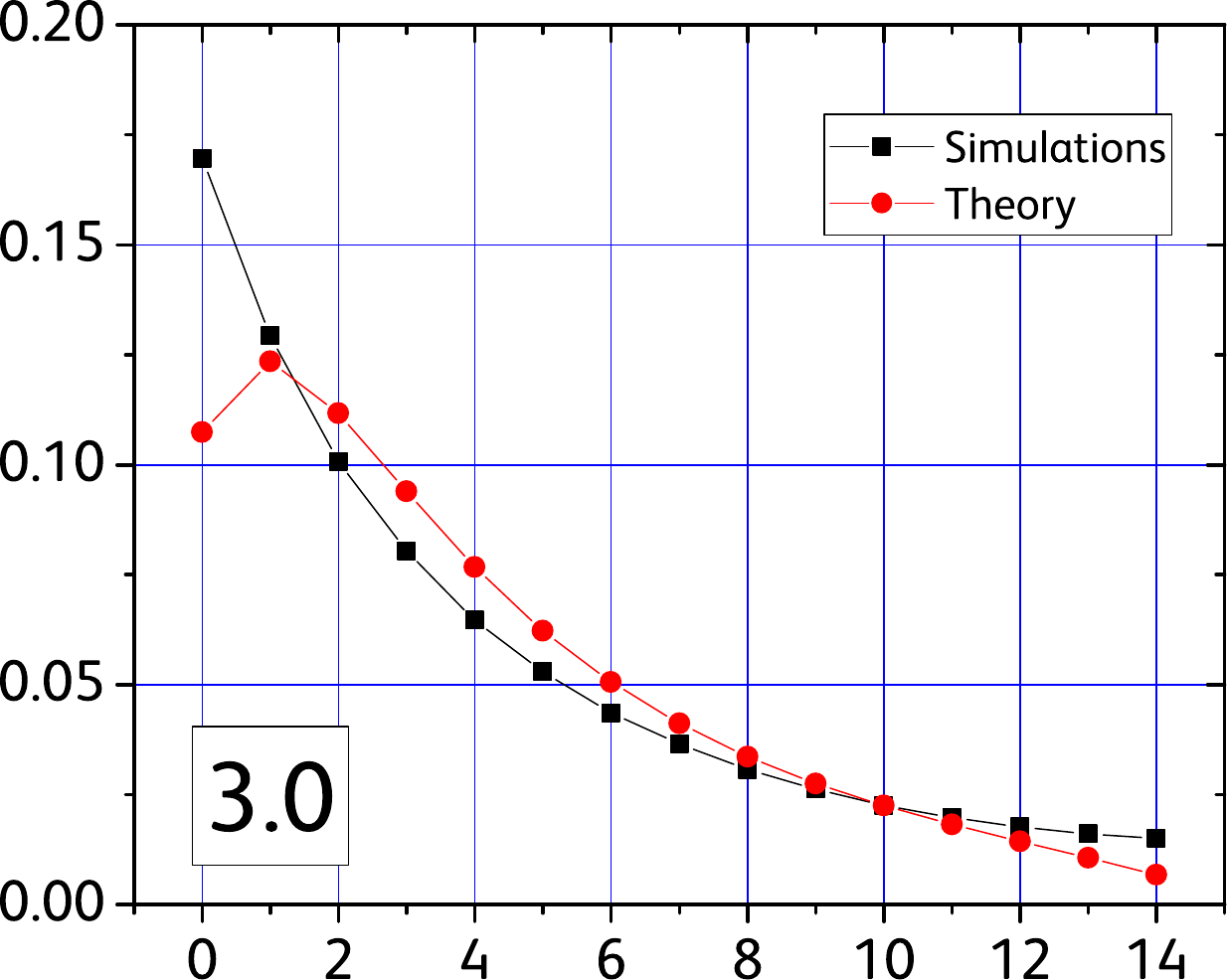}
\end{center}
\caption{\label{fig_N_distribs}
	Distribution of power among mode groups as a function of $N$ for 3 channels
	with fundamental-mode input and characterized by interaction strengths
	$|\Lambda_{00}L| = 0.1, 1.0$, and $3.0$, which is equivalent to Fried parameter
	values of 204~mm, 51.3~mm, and 26.6~mm respectively. Based on 120 lowest-order LG
	modes. The total power in these modes is 0.996, 0.95, and 0.8 respectively, while
	the rest is dissipated to even higher order modes.
}
\end{figure}
To validate the model, we calculated this matrix for the 120 lowest-order LG modes with
$w=40$~mm at a wavelength of $\lambda=850$~nm and turbulent scales of $l_0 = 1.0$~mm,
$L_0 = 1.0$~m, resulting in $\mathcal{I}_{00}=-5.57$. A relatively small $L_0$ was
deliberately chosen to minimize the simulation errors associated with the numerical generation
of phase screens by means of the Fourier method. We scaled the matrix to achieve
the necessary interaction strengths $\Lambda_{00}L$ and calculated the matrix
exponent~(\ref{eq_matrix_exponent_turb}) applied to the initial vector $\mathbf{v}_0$ with
unit power in the fundamental mode and zero elsewhere.

As a reference, we used simulation results that were obtained by application of 10600
independent turbulence phase screens to the fundamental mode and calculating the overap with 120
lowest-order HG and LG modes. The strength of the turbulence was chosen by setting the
Fried parameter $r_0$ to a value that produces the same $\Lambda_{00}L$ as in the model
using~(\ref{eq_fried_for_strength}). The turbulence scales $l_0$ and $L_0$
were the same as for the model. Phase screen generation is performed based
on algorithm~\cite{B97} with 828 frequency components for each axis, with the lowest spatial
frequency (and the step between adjacent ones) of $f_0 = 0.2$~m$^{-1}$.
The simulation grid size is $1024\times1024$ points spaced at 0.267~mm.

To represent the distribution of power among modes in a concise way and to make the result independent
of the choice of LG vs. HG, we calculated the total power for groups of modes with the same
$N$. The results for the interaction strengths of 0.1, 1 and 3 are shown in
Fig.~\ref{fig_N_distribs}. It must be noted that the simulation results based on the HG and LG
mode families, after grouping by $N$, only differ by less than 2\% relative error, which
matches the expected averaging error.

As one can see at the interaction strength of 0.1 there is an excellent agreement with the
simulation results with the only exception at the largest $N$, where the limited size of
the matrix $\mathbf{\Lambda}$ does not take into account non-negligible scattering from
modes with $N>14$. At strength 1, the agreement becomes somewhat poorer, but still
the model represents the simulation results with rather high precision. At the strength of
3 the dominating error is observed for the fundamental mode, where the model significantly
underestimates its population. The main cause of this mismatch is the coherence between
modes, which is discussed in detail in the next section.

\section{Applicability of the model and discussion}\label{sec_turb_applicability}

\subsection{Model applicability}
The only simplification made in our framework is to neglect the optical phases of the spatial
modes. This does not affect the result if a single spatial mode is present in
the channel. Therefore, if only a single mode is excited or if it strongly dominates over all
other spatial modes, the analysis developed here is exact. However, if the original signal is a
coherent superposition of two or more modes, the given description becomes an approximation, as the two
modes have certain phase relations and cannot be accurately described within the chosen
optical field representation.

In fact, the amount of the transferred power is phase-dependent as the induced wave
may interfere with the power already in the given spatial mode, either constructively or
destructively. Therefore, the prediction for a coherent superposition is only fair for
certain circumstances, where phase averaging does not become significantly biased due to
the interference.
If different spatial modes are mutually incoherent, our analysis is certainly applicable,
as the two waves do not interfere. 

This limitation also has consequences for the case of the Gaussian beam propagation
studied. Although the initial beam comprises a single spatial mode, other spatial
modes are generated due to the turbulent process. The amplitudes of such generation process
may have a dependence on the phase, which results in phase correlations between different
modes. This breaks down assumptions and leads to prediction errors. Nevertheless, as
one can see, the model still has a substantial predicting power even for quite strong
turbulence levels.

Moreover, the original simulation did not include light propagation, i.e., technically, the phase
screens had zero thickness. When propagation over distances comparable to $z_R$ is
taken into account, relative phases between modes with different $N$ get smeared because
of the propagation-dependent Gouy phase. This results in even better agreement with our model.

\begin{figure}[!htbp]
\begin{center}
\includegraphics[width=\columnwidth]{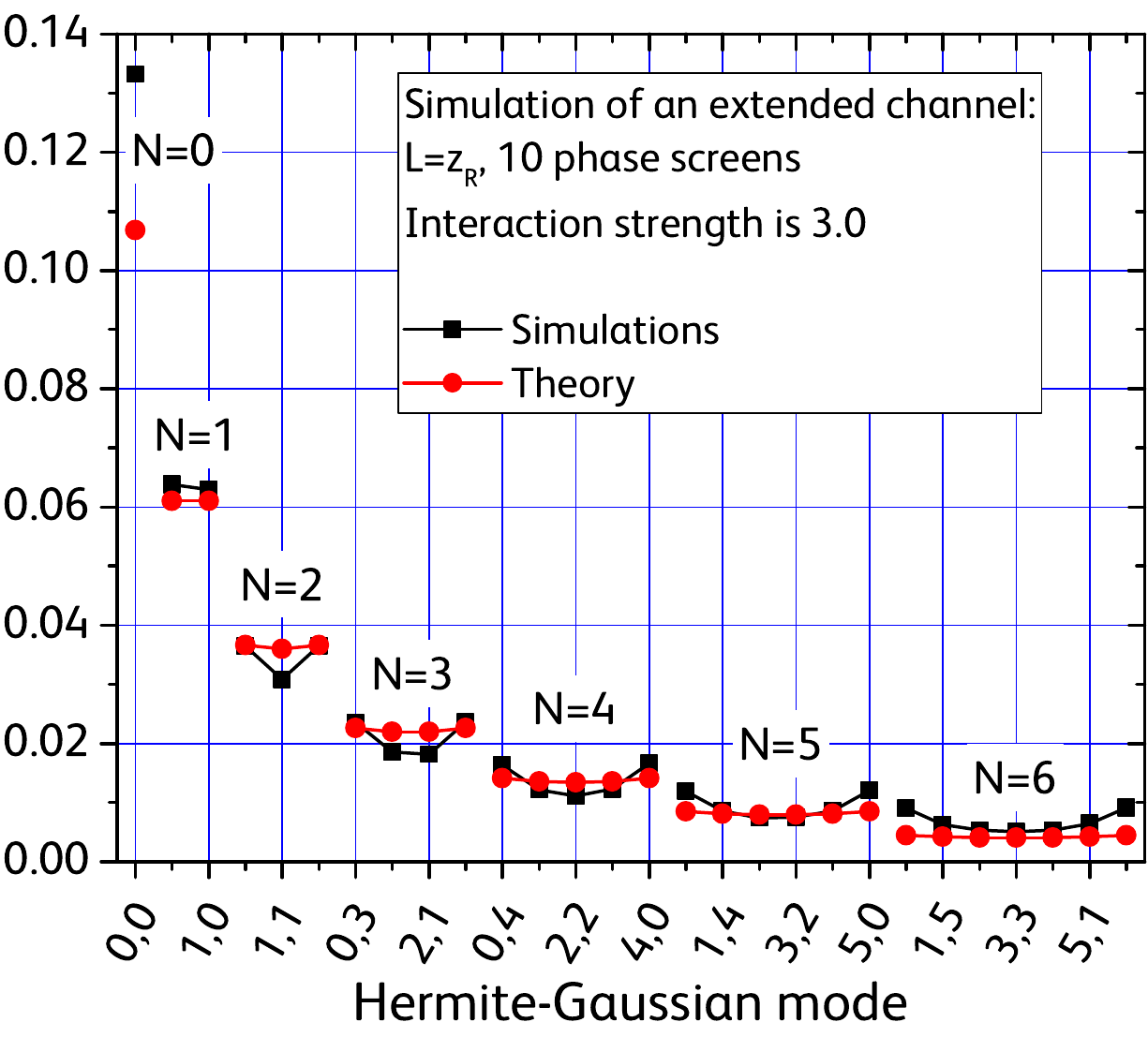}
\end{center}
\caption{\label{fig_propag_comparison}
	Distribution of power among HG modes in an extended channel with the length of
	$L=5900$~m and the interaction strength $|\Lambda_{00}L| = 3.0$. The Rytov parameter
	$\sigma_R^2 = 3.2$. Theoretical
	prediction, which does not depend on the channel length unless the beam diameter
	varies, is based on 28 lowest-order HG modes. Modes are presented in the following
	order: (0,0), (0,1), (1,0), (0,2), (1,1), (2,0), etc.
}
\end{figure}

As an example, we simulated the same problem, but included propagation between $-z_R/2$
and $z_R/2$ relative to the beam waist position. The total distance of 5900~m was divided
into 10
equal chunks with equal turbulence strengths in each. That makes the Rytov parameter for
the interscreen distance about 0.05. It ensures a good precision of simulation
because this value is below the widely-used margin of 0.1~\cite{KS23}.
Under these conditions, the radius $w$
of the spatial modes varied by less than 12\%, which was ignored in the theoretical model.
Figure~\ref{fig_propag_comparison} shows the results.

As one can see, in this extended channel the simulation result for the power in the
fundamental mode (0.133) is much closer to the theory prediction of 0.107 than the
previous simulation with no propagation (0.170). This is mainly due to the partial phase
randomization between different modes due to beam propagation. In general, there is strong
agreement between our theoretical model and the simulation, with notable deviation for
highest-order modes ($N=6$). For them, the theory prediction is again noticeably lower
due to the lack of power backflow from higher-order modes ignored while taking the matrix
exponent.

To validate our
simulation routine, we also simulated the same problem with 10 phase screens, but skipping
beam propagation steps. The results perfectly matched the original single phase screen simulation up to the averaging error of less than a few percent.

\subsection{Applicable turbulence strength}
The strength of turbulence in a channel is typically defined by the Rytov
parameter $\sigma_R^2$. Comparing~(\ref{eq_rytov_definition})
with~(\ref{eq_full_vk_integral}) we can express the Rytov parameter via the interaction
strength $|\Lambda_{00} L|$:
\begin{equation}
	\sigma_R^2 = 10.7\, \frac{\Lambda_{00}L}{\mathcal{I}_{00}}
	\left(\frac{L}{k w^2}\right)^{5/6}.
\end{equation}

In our simulation $|\mathcal{I}_{00}| = 5.57$, while it is upper-bounded by 11.9 (see the
next subsection). If we assume that the diameter of the beam does not change significantly
and $w(z) \approx w_0$, we can see that the ratio in parentheses is $L/(2z_R)$, which
should not be larger than 1 to preserve the same diameter of the mode along the
channel. Thus, in our example, the Rytov parameter is not larger than $1.9
|\Lambda_{00}L|$. For the strongest interaction case studied of $|\Lambda_{00}L| = 3$, the Rytov
parameter may be as large as 5.7. So we may say that our model is still reasonably
applicable even for the medium-to-strong turbulence regime.

\subsection{Convergence at $K\rightarrow 0$ and scalings for pure Kolmogorov
turbulence}\label{sec_diagonal_scaling}
One major problem associated with a description of turbulent processes is the divergence
of the Kolmogorov spectrum at $K\rightarrow 0$. Despite the available option of damping
low-frequency components, we show that the developed framework is fully compatible with
the undamped original spectrum~(\ref{eq_kolm_spec_n}). All acceptance spectra
$B_{ab}(\vartheta)$ vanish at $\vartheta = 0$ and behave at most linearly at small
arguments. This linear dependence only appears either for the same mode ($B_{aa} \approx
-2[N+1]\vartheta$) or for the case when $N$ changes by no more than 1. In all other cases,
the acceptance spectrum scales as a higher power of $\vartheta$.
Thus, the integral~(\ref{eq_lambda_definition}) always converges at zero even for
the Kolmogorov spectrum, in the worst case as $\int_0 \vartheta^{(-5/6)}\,d\vartheta$.
Its convergence at infinity is also guaranteed by the exponential decay of the
acceptance spectrum.
Therefore, there is no technical limitation to applying our model to the unmodified
Kolmogorov spectrum. Specifically, for the power in the fundamental mode, we have
\begin{multline}
	|\mathcal{I}_{00}^{\mathrm{max}}| = \int_0^\infty \vartheta^{-11/6}
	\bigl[1-\exp(-2\vartheta)\bigr]\,d\vartheta \\
	= \frac 65\, 2^{5/6}\Gamma(1/6)  \approx 11.902,
\end{multline}
which provides an upper bound for $\mathcal{I}_{00}$ for any damped
Kolmogorov spectrum.
Despite that this is technically feasible, it may not have much practical importance as this
marginal convergence at zero still results in a disproportionally large contribution from
$K\rightarrow 0$.

Nevertheless, the research of the behavior of various modes in pure Kolmogorov turbulence
draws enough attention~\cite{HRM13,LSB15,BSB19,BKI24}. This behavior can also be studied in the developed framework.
We are going to specifically examine the diagonal elements of the matrix
$\mathbf{\Lambda}$. As a recap, we emphasize that these diagonal elements are the exact
relative rates of power loss from a given mode with propagation distance, provided it is
the only mode excited in the channel. Thus, they indicate the strength of interaction
between the mode in question and the turbulent medium. As we show below, they have a very
prominent connection with the mode order $N$, a connection defined by the Kolmogorov spectral
shape.

Among the different modes, the diagonal elements $\Lambda_{aa}$ only differ by the value of the
dimensionless integral $\mathcal{I}_{aa}$, which will be studied here. Both expressions
for $B_{aa}(\vartheta)$ for the studied LG and HG mode families contain Laguerre
polynomials. Thus, we can use the Mehler–Heine formula, which describes their scaling with
increasing polynomial order:
\begin{equation}
	\lim_{n\rightarrow\infty} L_n(x) = J_0\left(2\sqrt{nx}\right).
\end{equation}
We start with the expression for LG modes, where we replace $\vartheta$ with $x/N$ and $p$
with $\beta N$:
\begin{multline}
	\mathcal{I}_{pl;pl} = N^{5/6}
	\int_0^\infty x^{-11/6}\\
	\times \biggl(1-{\Bigl[L_{(1-\beta)N}\left(\tfrac xN\right)
	L_{\beta N}\left(\tfrac xN\right)\Bigr]}^2 \exp\left(-\tfrac
	{2x}{N}\right)\biggr)\,dx.
\end{multline}

For increasing $N$ the exponential term becomes unity, while Laguerre polynomials may
be converted to Bessel functions. At this step, it becomes apparent that
the dependence on $N$ disappears within the integral. Moreover, since the $x^{-11/6}$ term
gives a huge preference to small values of $x$, if we substitute $J_0(2t)\approx1-t^2$ we can
see that the dependence on $\beta$ also disappears. In reality, it exists, but is expected
to be small.

Similar calculations for HG modes can be done with defining $\beta$ as $m/N$. We have
\begin{multline}
	\lim_{N\rightarrow\infty} \mathcal{I}_{mn;mn} = N^{5/6}
	\int_0^\infty x^{-11/6}\\
	\times \biggl(1-{\Bigl[J_0\left(2\cos\xi \sqrt{2\beta x}\right)
	J_0\left(2\sin\xi\sqrt{2(1-\beta)x}\right)\Bigr]}^2\biggr)\,dx.
\end{multline}
If, in addition, we use the approximation for small arguments of $J_0$ and perform averaging over
$\xi$, we also see that the dependence on $\beta$ disappears.

Therefore, for both mode families we see a
clean asymptotics $\mathcal{I}\sim N^{5/6}$ with some minor dependence on the particular
mode of order $N$. Explicit calculations for small mode orders reveal that the scaling
is, in fact, $\mathcal{I} \approx 12{(N+1)}^{5/6}$, which is illustrated in
Table~\ref{tab_Inm_values}. It contains both the exact values of $\mathcal{I}$ and this
simplified scaling. As one can see, the
relative error of such an approximation for the given set of modes is below 3\% for HG modes
and below 1\% for LG modes!

\begin{table*}
	\setlength{\tabcolsep}{5.5pt}
	\begin{tabular}{ccccccccccccc}
		$N$ 		& 0 	& 1 	& 2 	& 2 	& 3	& 3	& 4	&
		4	& 4 & 5 & 5 & 5\\
		\midrule
		LG	& 	& (0,-1)& (0,-2)&	& (0,-3)& (1,-1)& (0,-4)&
				(1,-2)& &(0,-5) & (1,-3) & (2,-1)\\
		modes 	& (0,0) & (0,1)	& (0,2)& (1,0)	& (0,3) & (1,1)	& (0,4)&
		(1,2)&	(2,0) & (0,5) & (1,3) & (2,1)\\
		$\mathcal{I}_{pl;pl}$ & 11.902& 21.406& 30.088& 29.961& 38.281& 38.130&
		46.130& 45.981& 45.892& 53.716& 53.577& 53.450\\
		\midrule
		HG	& 	& (0,1)& (0,2)&	& (0,3)& (1,2)& (0,4)&
				(1,3)& &(0,5) & (1,4) & (2,3)\\
		modes 	& (0,0) & (1,0)	& (2,0)& (1,1)	& (3,0) & (2,1)	& (4,0)&
		(3,1)&	(2,2) & (5,0) & (4,1) & (3,2)\\
		$\mathcal{I}_{mn;mn}$ & 11.902& 21.200& 29.558& 30.056& 37.411& 38.145&
		44.922& 45.801& 46.054& 52.174& 53.155& 53.571\\

		\midrule
		$12{(N+1)}^{5/6}$ & 12.000 & 21.382 & 29.977 & 29.977 & 38.098 & 38.098 &
		45.883& 45.883& 45.883& 53.412& 53.412& 53.412\\
	\end{tabular}
	\caption{\label{tab_Inm_values}
	Exact values of the dimensionless integral $\mathcal{I}_{aa}$ for lowest order
	LG and HG modes and their approximation with a function of $N$. These
	values define the relative rates of power loss from a given mode with propagation
	distance for Kolmogorov turbulence.}
\end{table*}

Hence, we have shown that the strength of mode interaction with a turbulent medium
scales as ${(N+1)}^{5/6}$. Using (\ref{eq_fried_for_strength}) one can see that to achieve
the same level of power dissipation $\Lambda_{aa}L$, the value of ${(w/r_0)}^{5/3}$ should
be inversely proportional to $\mathcal{I}_{aa}$. That translates to $w\sqrt{N+1}/r_0 =
\mathrm{const.}$, which corresponds to a universal scaling law for the interaction
strength with respect to the mode order. It becomes even more intuitive if one recalls
that the intensity-weighted RMS
radius of the beam for both HG and LG modes is $w\sqrt{(N+1)/2}$. So, the relative amount of
power loss for different modes is only a function of the ratio of their RMS size and the Fried
parameter.

The very same scaling for LG modes was previously empirically found in~\cite{HRM13}, later confirmed
in~\cite{LSB15,BKI24}, and justified through asymptotic analysis~\cite{BSB19}. The authors of these works observe a universal dependence of
some measurable quantity (like the concurrence or the ratio of the received power in modes
with positive and negative values of $l$) for different modes, given that the Fried
parameter is scaled by what they call the {\em phase correlation length}~\cite{LSB15}. Indeed, these
measurable parameters are directly tied to the strength of interaction with the turbulent
medium, which scales with the mode order, as we have just demonstrated.

The only missing part in observing that the two scalings are indeed the same is
a connection between the phase correlation length $\xi$ and the mode order. Their
definition of $\xi$ for the mode $LG_{pl}$ is $\xi=\sin[\pi/(2|l|)]\langle\rho\rangle$,
where $\langle\rho\rangle$ is the intensity-weighted mean radius of the mode. This can be
fairly accurately connected to the mentioned earlier RMS radius, because their largest
mismatch is observed for the fundamental mode and is only around 13\%, while it is
significantly lower for higher-order modes with low values of $p$.

Papers~\cite{HRM13,LSB15,BSB19} only consider the case with $p=0$, therefore $N=|l|$ and, thus,
replacing the sine function with its argument, we get the scaling $\xi \approx 1/\sqrt{N}$. It
can be easily checked that for $|l|>1$ this is a consistent approximation, which matches
our results. In~\cite{BKI24} the authors also considered the case of nonzero $p$, however, as
one can see from figure SF4, for $p=5$, unlike the case $p=0$, there is a significant separation
between small and large values of $|l|$. This is because the real scaling of the
interaction strength is given by the RMS mode size, while the phase correlation length
was an empirical estimate that becomes increasingly inaccurate for larger values of $p$.

\subsection{Applications for adaptive pointing systems}

Another observation, which may become very handy in practical applications, comes from the
flexibility of our framework to accommodate arbitrary turbulence spectra. Most practical
free-space optical links use some kind of a pointing, acquisition, and tracking (PAT)
system. They significantly help in reduction of the channel loss in the presence of
turbulence.
A more advanced version of the PAT is some kind of adaptive optics for further correction of
turbulent distortions.

All of these tools may be viewed as high-pass spatial filters for turbulent distortions,
which effectively suppress phase fluctuations at low spatial frequencies, perceived in the
first place as the angle of arrival fluctuations. Measuring the spectral response of these
tools, one can find the effective residual phase fluctuations spectrum observed in the
corrected optical channel. This residual spectrum can then be used within our framework to
predict channel performance. Such an additional damping of the turbulence
spectrum results in weaker interactions between spatial modes and, consequently, the improved
overall performance.

\subsection{Comparison with previous results}

There are a very limited number of previous publications in which optical power in a
particular spatial mode was studied. One notable example is~\cite{TB09}, which analyzed the
crosstalk for the pure phase vortex beams having uniform intensity within a circle.
Although they are substantially different from the LG modes, the result for the power
remaining in the same mode is $1-1.01(D/r_0)^{5/3}$ at the limit of $r_0 \rightarrow
\infty$, where $D$ is the
diameter of the beam.  Using~(\ref{eq_fried_definition}) one can see that it indicates 
linear scaling with $L$, which exactly matches our findings for small perturbations.
Similarly, the power in adjacent modes with a different OAM is shown to increase
linearly, which also agrees with the present analysis. Unfortunately, the results for high levels of
turbulence do not show a good agreement both because our model becomes less accurate and
because the pure vortex beams are not LG modes studied in our framework.

A similar picture appears when comparing with~\cite{P05}. Its authors studied families of
spatial modes carrying the same OAM.
In fact, they analyzed a mode set, where the radial part is independent of the azimuthal
mode order. This is not the case for LG modes, where the radial part depends on both
the radial $p$ and the azimuthal $l$ parameters. Neither are they eigenfunctions of
propagation, which is a critical requirement of our model. Thus, we can only confirm that the
behavior under low turbulence levels shows a reasonably good agreement with this previous
work. The matching is the best when more than 50\% of the power remains with the particular
OAM value. At even stronger turbulence levels, the two models become too different for a direct
comparison.

A more accurate comparison can be made with the first-order approximation for the Gaussian
beam from our previous work~\cite{KZR18}. The strength of the turbulent action $|P_{00}|$,
denoted $w^2 C_a/2$ in the cited work, is given by the same
integral~(\ref{eq_pab_action_strength}), but with the filtering function of $2\vartheta
|F(K)|^2$ instead of $|B_{00}(\vartheta)| = 1-\exp(-2\vartheta)$, where the low-pass filter
$|F(K)|^2$ was introduced to account for the damping of high-frequency components due to
spatial averaging. Both expressions exhibit exactly the same asymptotic behavior as
$\vartheta\rightarrow 0$. The mean remaining power in the fundamental mode was
$(1+|P_{00}|)^{-1}$, whereas our present model, applied exclusively
to the fundamental mode, yields $\exp(-|P_{00}|)$. The two expressions essentially
coincide for small arguments and behave similarly for larger ones. Moreover, the
same scaling~(\ref{eq_bzerononzero}) with $N$ appears for scattering into higher-order modes.
Overall, there is excellent agreement with this previous result in the low-frequency
part of the spectrum and for not too strong turbulence, which is naturally expected given
the limitations of the first-order approximation.

\section{Conclusion}

In this paper, we presented a general approach describing spatial mode interactions induced
by atmospheric turbulence in a free-space optical channel. We developed a straightforward
framework and applied it to two common families of spatial modes: Hermite-Gaussian for
Cartesian symmetry and Laguerre-Gaussian for cylindrical symmetry.
The framework is based on the split-step propagation model, where we represent an optical
field as a discrete distribution of optical power among a complete set of spatial modes.
As this representation does not change with unperturbed propagation, the whole process
effectively becomes an integral perturbation induced by infinitesimal phase screens.

One of the obtained fundamental results is that the mode interaction strength is proportional
to the turbulent phase fluctuation power that overlaps with the spatial acceptance
spectrum, specific for a given pair of interacting modes. Similarly, the optical power lost from a
mode is also proportional to the total power of phase fluctuations within the specific
spatial spectrum. Thus, as the power of phase fluctuations accumulates additively along
the propagation, the induced interactions of the spatial modes are proportional to the
propagation distance in uniform channels.

This result is exact for channels with a single excited spatial mode (or an incoherent
combination of them) and weak turbulent perturbations. Taking into account a larger number
of spatial modes for calculation of their dynamics, the same framework can be used to
reliably predict the average distribution of optical power among spatial modes under
medium-to-strong turbulence conditions.
Comparison with full-scale simulations shows that the accuracy of our model improves with beam propagation for a given phase fluctuation power,
as propagation smears phase relations between different modes. The matching would be ideal if the phases
were randomized at each propagation step, which aligns with the assumption of full
incoherence.

The developed framework helps to better understand the behavior of structured light in
turbulent propagation conditions and it can also be used for quantitative estimations
of modal crosstalk and the number of excited modes. The analytical expressions are also
suitable for validation and debugging of various simulation tools especially for unbounded
spectra like Kolmogorov's, whose accurate accounting in simulations is
nontrivial~\cite{PCF25}. The developed tools were used to demonstrate the existence
of a universal connection between the mode number and the rate of power dissipation, which
was earlier found empirically for a subset of LG modes.

A particular advantage of the proposed model is its spectral-agnostic nature. It can be
readily applied to the Kolmogorov spectrum and its derivatives, as well as to arbitrary
spectra, which may appear in other types of stochastic propagation media.
Similarly, it can model the behavior of a channel that is equipped with a precision optical
tracking and adaptive optic systems. Those may be perceived
as equivalent high-pass filters for the turbulent process, which effectively reduce the
power of phase fluctuations and suppress modal crosstalk.

\section*{Acknowledgment}
The author thanks James Grieve and Sana Amairi-Pyka for their support and assistance with
the manuscript, and Stephen Vintskevich for stimulating discussions.

\section*{Appendix}
A summary of the key equations is given below. The main variable $\vartheta$ is defined as
$\vartheta = K^2 w^2/8$.
\subsection{Acceptance filters for interacting mode pairs}
For LG modes ($q,s \ne p,l$):
\begin{multline}
	B_{pl;qs}^{LG}(\vartheta) =
	\frac{q!}{(q+|s|)!}
	{\vartheta}^{|s|} \exp(-2\vartheta)\\
	\times\begin{cases}
		\frac{(p+|l|)!}{p!} {\vartheta}^{-|l|}
		\left[
			L_{p+|l|}^{q-p+|s|-|l|}(\vartheta)
			L_q^{p-q}(\vartheta)
			\right]^2 &\text{if}\; ls\ge 0,\\
		\frac{p!}{(p+|l|)!} {\vartheta}^{|l|}
		\left[
			L_p^{q-p+|s|}(\vartheta)
			L_q^{p-q+|l|}(\vartheta)
			\right]^2 &\text{if}\; ls\le 0.
	\end{cases}
\end{multline}
\begin{equation}
	B_{pl;pl}^{LG}(\vartheta) = 
	{\left[L_{p+|l|}(\vartheta)L_p(\vartheta)\right]}^2
	\exp(-2\vartheta) - 1.
\end{equation}

For HG modes (requires additional averaging over the direction $\xi\in[0,\pi/2]$; $k,t\ne m,n$):
\begin{multline}
	B_{mn;kt}^{HG}(\vartheta) =
	\frac{m!\,n!}{k!\,t!}
	{\bigl(2\vartheta\cos^2\xi\bigr)}^{k-m}
	{\bigl(2\vartheta\sin^2\xi\bigr)}^{t-n}\\
	\times{\Bigl[
		L_m^{k-m}\bigl(2\vartheta\cos^2\xi\bigr)
		L_n^{t-n}\bigl(2\vartheta\sin^2\xi\bigr)
		\Bigr]}^2 \exp(-2\vartheta)
\end{multline}
(requires ordering of $m$ and $k$ such that $m\le k$, ordering $n$ and $t$ such that $n\le
t$)
\begin{multline}
	B_{mn;mn}^{HG}(\vartheta) = \\
	{\Bigl(
	L_m\bigl(2\vartheta\cos^2\xi\bigr)
	L_n\bigl(2\vartheta\sin^2\xi\bigr)\Bigr)}^2 \exp( -2\vartheta) - 1.
\end{multline}

\subsection{Spectral overlap integrals}
For the modified von Karman spectrum with the parameters $l_0$ and $L_0$
\begin{multline}
\mathcal{I}_{ab} = \int_0^\infty {\left(\vartheta+\frac{\pi^2 w^2}{2 L_0^2}\right)}^{-11/6}\\
\times\exp\Bigl(-2\vartheta\, l_0^2 /(\pi^2w^2)\Bigr) \,B_{ab}(\vartheta)\,d\vartheta.
\end{multline}
For the Kolmogorov spectrum set $l_0=L_0^{-1} = 0$.

\subsection{Mode connections matrix}
For a given structure constant:
\begin{equation}
\Lambda_{ab} = 0.115 C_n^2 k^2 w^{5/3} \mathcal{I}_{ab}
\end{equation}
For a given Fried parameter:
\begin{equation}
 \Lambda_{ab} = 0.272 \left(\frac{w}{r_0}\right)^{5/3} \frac{\mathcal{I}_{ab}}{L}
\end{equation}


\begin{thebibliography}{37}%
\makeatletter
\providecommand \@ifxundefined [1]{%
 \@ifx{#1\undefined}
}%
\providecommand \@ifnum [1]{%
 \ifnum #1\expandafter \@firstoftwo
 \else \expandafter \@secondoftwo
 \fi
}%
\providecommand \@ifx [1]{%
 \ifx #1\expandafter \@firstoftwo
 \else \expandafter \@secondoftwo
 \fi
}%
\providecommand \natexlab [1]{#1}%
\providecommand \enquote  [1]{``#1''}%
\providecommand \bibnamefont  [1]{#1}%
\providecommand \bibfnamefont [1]{#1}%
\providecommand \citenamefont [1]{#1}%
\providecommand \href@noop [0]{\@secondoftwo}%
\providecommand \href [0]{\begingroup \@sanitize@url \@href}%
\providecommand \@href[1]{\@@startlink{#1}\@@href}%
\providecommand \@@href[1]{\endgroup#1\@@endlink}%
\providecommand \@sanitize@url [0]{\catcode `\\12\catcode `\$12\catcode
  `\&12\catcode `\#12\catcode `\^12\catcode `\_12\catcode `\%12\relax}%
\providecommand \@@startlink[1]{}%
\providecommand \@@endlink[0]{}%
\providecommand \url  [0]{\begingroup\@sanitize@url \@url }%
\providecommand \@url [1]{\endgroup\@href {#1}{\urlprefix }}%
\providecommand \urlprefix  [0]{URL }%
\providecommand \Eprint [0]{\href }%
\providecommand \doibase [0]{https://doi.org/}%
\providecommand \selectlanguage [0]{\@gobble}%
\providecommand \bibinfo  [0]{\@secondoftwo}%
\providecommand \bibfield  [0]{\@secondoftwo}%
\providecommand \translation [1]{[#1]}%
\providecommand \BibitemOpen [0]{}%
\providecommand \bibitemStop [0]{}%
\providecommand \bibitemNoStop [0]{.\EOS\space}%
\providecommand \EOS [0]{\spacefactor3000\relax}%
\providecommand \BibitemShut  [1]{\csname bibitem#1\endcsname}%
\let\auto@bib@innerbib\@empty
\bibitem [{\citenamefont {Mirhosseini}\ \emph {et~al.}(2015)\citenamefont
  {Mirhosseini}, \citenamefont {Magana-Loaiza}, \citenamefont {O'Sullivan},
  \citenamefont {Rodenburg}, \citenamefont {Malik}, \citenamefont {Lavery},
  \citenamefont {Padgett}, \citenamefont {Gauthier},\ and\ \citenamefont
  {Boyd}}]{MMS15}%
  \BibitemOpen
  \bibfield  {author} {\bibinfo {author} {\bibfnamefont {M.}~\bibnamefont
  {Mirhosseini}}, \bibinfo {author} {\bibfnamefont {O.~S.}\ \bibnamefont
  {Magana-Loaiza}}, \bibinfo {author} {\bibfnamefont {M.~N.}\ \bibnamefont
  {O'Sullivan}}, \bibinfo {author} {\bibfnamefont {B.}~\bibnamefont
  {Rodenburg}}, \bibinfo {author} {\bibfnamefont {M.}~\bibnamefont {Malik}},
  \bibinfo {author} {\bibfnamefont {M.~P.~J.}\ \bibnamefont {Lavery}}, \bibinfo
  {author} {\bibfnamefont {M.~J.}\ \bibnamefont {Padgett}}, \bibinfo {author}
  {\bibfnamefont {D.~J.}\ \bibnamefont {Gauthier}},\ and\ \bibinfo {author}
  {\bibfnamefont {R.~W.}\ \bibnamefont {Boyd}},\ }\bibfield  {title} {\bibinfo
  {title} {High-dimensional quantum cryptography with twisted light},\ }\href
  {https://doi.org/10.1088/1367-2630/17/3/033033} {\bibfield  {journal}
  {\bibinfo  {journal} {New J. Phys.}\ }\textbf {\bibinfo {volume} {17}},\
  \bibinfo {pages} {033033} (\bibinfo {year} {2015})}\BibitemShut {NoStop}%
\bibitem [{\citenamefont {Sit}\ \emph {et~al.}(2017)\citenamefont {Sit},
  \citenamefont {Bouchard}, \citenamefont {Fickler}, \citenamefont
  {Gagnon-Bischoff}, \citenamefont {Larocque}, \citenamefont {Heshami},
  \citenamefont {Elser}, \citenamefont {Peuntinger}, \citenamefont {Gunthner},
  \citenamefont {Heim}, \citenamefont {Marquardt}, \citenamefont {Leuchs},
  \citenamefont {Boyd},\ and\ \citenamefont {Karimi}}]{SBF17}%
  \BibitemOpen
  \bibfield  {author} {\bibinfo {author} {\bibfnamefont {A.}~\bibnamefont
  {Sit}}, \bibinfo {author} {\bibfnamefont {F.}~\bibnamefont {Bouchard}},
  \bibinfo {author} {\bibfnamefont {R.}~\bibnamefont {Fickler}}, \bibinfo
  {author} {\bibfnamefont {J.}~\bibnamefont {Gagnon-Bischoff}}, \bibinfo
  {author} {\bibfnamefont {H.}~\bibnamefont {Larocque}}, \bibinfo {author}
  {\bibfnamefont {K.}~\bibnamefont {Heshami}}, \bibinfo {author} {\bibfnamefont
  {D.}~\bibnamefont {Elser}}, \bibinfo {author} {\bibfnamefont
  {C.}~\bibnamefont {Peuntinger}}, \bibinfo {author} {\bibfnamefont
  {K.}~\bibnamefont {Gunthner}}, \bibinfo {author} {\bibfnamefont
  {B.}~\bibnamefont {Heim}}, \bibinfo {author} {\bibfnamefont {C.}~\bibnamefont
  {Marquardt}}, \bibinfo {author} {\bibfnamefont {G.}~\bibnamefont {Leuchs}},
  \bibinfo {author} {\bibfnamefont {R.~W.}\ \bibnamefont {Boyd}},\ and\
  \bibinfo {author} {\bibfnamefont {E.}~\bibnamefont {Karimi}},\ }\bibfield
  {title} {\bibinfo {title} {High-dimensional intracity quantum cryptography
  with structured photons},\ }\href {https://doi.org/10.1364/OPTICA.4.001006}
  {\bibfield  {journal} {\bibinfo  {journal} {Optica}\ }\textbf {\bibinfo
  {volume} {4}},\ \bibinfo {pages} {1006} (\bibinfo {year} {2017})}\BibitemShut
  {NoStop}%
\bibitem [{\citenamefont {Cozzolino}\ \emph {et~al.}(2019)\citenamefont
  {Cozzolino}, \citenamefont {Da~Lio}, \citenamefont {Bacco},\ and\
  \citenamefont {Oxenløwe}}]{CLB19}%
  \BibitemOpen
  \bibfield  {author} {\bibinfo {author} {\bibfnamefont {D.}~\bibnamefont
  {Cozzolino}}, \bibinfo {author} {\bibfnamefont {B.}~\bibnamefont {Da~Lio}},
  \bibinfo {author} {\bibfnamefont {D.}~\bibnamefont {Bacco}},\ and\ \bibinfo
  {author} {\bibfnamefont {L.~K.}\ \bibnamefont {Oxenløwe}},\ }\bibfield
  {title} {\bibinfo {title} {High-dimensional quantum communication: Benefits,
  progress, and future challenges},\ }\href
  {https://doi.org/10.1002/qute.201900038} {\bibfield  {journal} {\bibinfo
  {journal} {Advanced Quantum Technologies}\ }\textbf {\bibinfo {volume} {2}},\
  \bibinfo {pages} {1900038} (\bibinfo {year} {2019})}\BibitemShut {NoStop}%
\bibitem [{\citenamefont {Miller}(2000)}]{M00}%
  \BibitemOpen
  \bibfield  {author} {\bibinfo {author} {\bibfnamefont {D.~A.~B.}\
  \bibnamefont {Miller}},\ }\bibfield  {title} {\bibinfo {title} {Communicating
  with waves between volumes: evaluating orthogonal spatial channels and limits
  on coupling strengths},\ }\href {https://doi.org/10.1364/AO.39.001681}
  {\bibfield  {journal} {\bibinfo  {journal} {Appl. Opt.}\ }\textbf {\bibinfo
  {volume} {39}},\ \bibinfo {pages} {1681} (\bibinfo {year}
  {2000})}\BibitemShut {NoStop}%
\bibitem [{\citenamefont {Paterson}(2005)}]{P05}%
  \BibitemOpen
  \bibfield  {author} {\bibinfo {author} {\bibfnamefont {C.}~\bibnamefont
  {Paterson}},\ }\bibfield  {title} {\bibinfo {title} {Atmospheric turbulence
  and orbital angular momentum of single photons for optical communication},\
  }\href {https://doi.org/10.1103/PhysRevLett.94.153901} {\bibfield  {journal}
  {\bibinfo  {journal} {Phys. Rev. Lett.}\ }\textbf {\bibinfo {volume} {94}},\
  \bibinfo {pages} {153901} (\bibinfo {year} {2005})}\BibitemShut {NoStop}%
\bibitem [{\citenamefont {Tyler}\ and\ \citenamefont {Boyd}(2009)}]{TB09}%
  \BibitemOpen
  \bibfield  {author} {\bibinfo {author} {\bibfnamefont {G.~A.}\ \bibnamefont
  {Tyler}}\ and\ \bibinfo {author} {\bibfnamefont {R.~W.}\ \bibnamefont
  {Boyd}},\ }\bibfield  {title} {\bibinfo {title} {Influence of atmospheric
  turbulence on the propagation of quantum states of light carrying orbital
  angular momentum},\ }\href {https://doi.org/10.1364/OL.34.000142} {\bibfield
  {journal} {\bibinfo  {journal} {Opt. Lett.}\ }\textbf {\bibinfo {volume}
  {34}},\ \bibinfo {pages} {142} (\bibinfo {year} {2009})}\BibitemShut
  {NoStop}%
\bibitem [{\citenamefont {Rodenburg}\ \emph {et~al.}(2012)\citenamefont
  {Rodenburg}, \citenamefont {Lavery}, \citenamefont {Malik}, \citenamefont
  {O'Sullivan}, \citenamefont {Mirhosseini}, \citenamefont {Robertson},
  \citenamefont {Padgett},\ and\ \citenamefont {Boyd}}]{RLM12}%
  \BibitemOpen
  \bibfield  {author} {\bibinfo {author} {\bibfnamefont {B.}~\bibnamefont
  {Rodenburg}}, \bibinfo {author} {\bibfnamefont {M.~P.~J.}\ \bibnamefont
  {Lavery}}, \bibinfo {author} {\bibfnamefont {M.}~\bibnamefont {Malik}},
  \bibinfo {author} {\bibfnamefont {M.~N.}\ \bibnamefont {O'Sullivan}},
  \bibinfo {author} {\bibfnamefont {M.}~\bibnamefont {Mirhosseini}}, \bibinfo
  {author} {\bibfnamefont {D.~J.}\ \bibnamefont {Robertson}}, \bibinfo {author}
  {\bibfnamefont {M.}~\bibnamefont {Padgett}},\ and\ \bibinfo {author}
  {\bibfnamefont {R.~W.}\ \bibnamefont {Boyd}},\ }\bibfield  {title} {\bibinfo
  {title} {Influence of atmospheric turbulence on states of light carrying
  orbital angular momentum},\ }\href {https://doi.org/10.1364/OL.37.003735}
  {\bibfield  {journal} {\bibinfo  {journal} {Opt. Lett.}\ }\textbf {\bibinfo
  {volume} {37}},\ \bibinfo {pages} {3735} (\bibinfo {year}
  {2012})}\BibitemShut {NoStop}%
\bibitem [{\citenamefont {Bachmann}\ \emph {et~al.}(2024)\citenamefont
  {Bachmann}, \citenamefont {Klug}, \citenamefont {Isoard}, \citenamefont
  {Shatokhin}, \citenamefont {Sorelli}, \citenamefont {Buchleitner},\ and\
  \citenamefont {Forbes}}]{BKI24}%
  \BibitemOpen
  \bibfield  {author} {\bibinfo {author} {\bibfnamefont {D.}~\bibnamefont
  {Bachmann}}, \bibinfo {author} {\bibfnamefont {A.}~\bibnamefont {Klug}},
  \bibinfo {author} {\bibfnamefont {M.}~\bibnamefont {Isoard}}, \bibinfo
  {author} {\bibfnamefont {V.}~\bibnamefont {Shatokhin}}, \bibinfo {author}
  {\bibfnamefont {G.}~\bibnamefont {Sorelli}}, \bibinfo {author} {\bibfnamefont
  {A.}~\bibnamefont {Buchleitner}},\ and\ \bibinfo {author} {\bibfnamefont
  {A.}~\bibnamefont {Forbes}},\ }\bibfield  {title} {\bibinfo {title}
  {Universal crosstalk of twisted light in random media},\ }\href
  {https://doi.org/10.1103/PhysRevLett.132.063801} {\bibfield  {journal}
  {\bibinfo  {journal} {Phys. Rev. Lett.}\ }\textbf {\bibinfo {volume} {132}},\
  \bibinfo {pages} {063801} (\bibinfo {year} {2024})}\BibitemShut {NoStop}%
\bibitem [{\citenamefont {Hamadou~Ibrahim}\ \emph {et~al.}(2013)\citenamefont
  {Hamadou~Ibrahim}, \citenamefont {Roux}, \citenamefont {McLaren},
  \citenamefont {Konrad},\ and\ \citenamefont {Forbes}}]{HRM13}%
  \BibitemOpen
  \bibfield  {author} {\bibinfo {author} {\bibfnamefont {A.}~\bibnamefont
  {Hamadou~Ibrahim}}, \bibinfo {author} {\bibfnamefont {F.~S.}\ \bibnamefont
  {Roux}}, \bibinfo {author} {\bibfnamefont {M.}~\bibnamefont {McLaren}},
  \bibinfo {author} {\bibfnamefont {T.}~\bibnamefont {Konrad}},\ and\ \bibinfo
  {author} {\bibfnamefont {A.}~\bibnamefont {Forbes}},\ }\bibfield  {title}
  {\bibinfo {title} {Orbital-angular-momentum entanglement in turbulence},\
  }\href {https://doi.org/10.1103/PhysRevA.88.012312} {\bibfield  {journal}
  {\bibinfo  {journal} {Phys. Rev. A}\ }\textbf {\bibinfo {volume} {88}},\
  \bibinfo {pages} {012312} (\bibinfo {year} {2013})}\BibitemShut {NoStop}%
\bibitem [{\citenamefont {Leonhard}\ \emph {et~al.}(2015)\citenamefont
  {Leonhard}, \citenamefont {Shatokhin},\ and\ \citenamefont
  {Buchleitner}}]{LSB15}%
  \BibitemOpen
  \bibfield  {author} {\bibinfo {author} {\bibfnamefont {N.~D.}\ \bibnamefont
  {Leonhard}}, \bibinfo {author} {\bibfnamefont {V.~N.}\ \bibnamefont
  {Shatokhin}},\ and\ \bibinfo {author} {\bibfnamefont {A.}~\bibnamefont
  {Buchleitner}},\ }\bibfield  {title} {\bibinfo {title} {Universal
  entanglement decay of photonic-orbital-angular-momentum qubit states in
  atmospheric turbulence},\ }\href {https://doi.org/10.1103/PhysRevA.91.012345}
  {\bibfield  {journal} {\bibinfo  {journal} {Phys. Rev. A}\ }\textbf {\bibinfo
  {volume} {91}},\ \bibinfo {pages} {012345} (\bibinfo {year}
  {2015})}\BibitemShut {NoStop}%
\bibitem [{\citenamefont {Aksenov}\ \emph {et~al.}(2016)\citenamefont
  {Aksenov}, \citenamefont {Kolosov}, \citenamefont {Filimonov},\ and\
  \citenamefont {Pogutsa}}]{AKF16}%
  \BibitemOpen
  \bibfield  {author} {\bibinfo {author} {\bibfnamefont {V.~P.}\ \bibnamefont
  {Aksenov}}, \bibinfo {author} {\bibfnamefont {V.~V.}\ \bibnamefont
  {Kolosov}}, \bibinfo {author} {\bibfnamefont {G.~A.}\ \bibnamefont
  {Filimonov}},\ and\ \bibinfo {author} {\bibfnamefont {C.~E.}\ \bibnamefont
  {Pogutsa}},\ }\bibfield  {title} {\bibinfo {title} {Orbital angular momentum
  of a laser beam in a turbulent medium: preservation of the average value and
  variance of fluctuations},\ }\href
  {https://doi.org/10.1088/2040-8978/18/5/054013} {\bibfield  {journal}
  {\bibinfo  {journal} {Journal of Optics}\ }\textbf {\bibinfo {volume} {18}},\
  \bibinfo {pages} {054013} (\bibinfo {year} {2016})}\BibitemShut {NoStop}%
\bibitem [{\citenamefont {Ndagano}\ \emph {et~al.}(2017)\citenamefont
  {Ndagano}, \citenamefont {Mphuthi}, \citenamefont {Milione},\ and\
  \citenamefont {Forbes}}]{NMM17}%
  \BibitemOpen
  \bibfield  {author} {\bibinfo {author} {\bibfnamefont {B.}~\bibnamefont
  {Ndagano}}, \bibinfo {author} {\bibfnamefont {N.}~\bibnamefont {Mphuthi}},
  \bibinfo {author} {\bibfnamefont {G.}~\bibnamefont {Milione}},\ and\ \bibinfo
  {author} {\bibfnamefont {A.}~\bibnamefont {Forbes}},\ }\bibfield  {title}
  {\bibinfo {title} {Comparing mode-crosstalk and mode-dependent loss of
  laterally displaced orbital angular momentum and {H}ermite-{G}aussian modes
  for free-space optical communication},\ }\href
  {https://doi.org/10.1364/OL.42.004175} {\bibfield  {journal} {\bibinfo
  {journal} {Opt. Lett.}\ }\textbf {\bibinfo {volume} {42}},\ \bibinfo {pages}
  {4175} (\bibinfo {year} {2017})}\BibitemShut {NoStop}%
\bibitem [{\citenamefont {Cox}\ \emph {et~al.}(2021)\citenamefont {Cox},
  \citenamefont {Mphuthi}, \citenamefont {Nape}, \citenamefont {Mashaba},
  \citenamefont {Cheng},\ and\ \citenamefont {Forbes}}]{CMN21}%
  \BibitemOpen
  \bibfield  {author} {\bibinfo {author} {\bibfnamefont {M.~A.}\ \bibnamefont
  {Cox}}, \bibinfo {author} {\bibfnamefont {N.}~\bibnamefont {Mphuthi}},
  \bibinfo {author} {\bibfnamefont {I.}~\bibnamefont {Nape}}, \bibinfo {author}
  {\bibfnamefont {N.}~\bibnamefont {Mashaba}}, \bibinfo {author} {\bibfnamefont
  {L.}~\bibnamefont {Cheng}},\ and\ \bibinfo {author} {\bibfnamefont
  {A.}~\bibnamefont {Forbes}},\ }\bibfield  {title} {\bibinfo {title}
  {Structured light in turbulence},\ }\href
  {https://doi.org/10.1109/JSTQE.2020.3023790} {\bibfield  {journal} {\bibinfo
  {journal} {IEEE J. Sel. Top. Quantum Electron.}\ }\textbf {\bibinfo {volume}
  {27}},\ \bibinfo {pages} {1} (\bibinfo {year} {2021})}\BibitemShut {NoStop}%
\bibitem [{\citenamefont {Peters}\ \emph {et~al.}(2025)\citenamefont {Peters},
  \citenamefont {Cocotos},\ and\ \citenamefont {Forbes}}]{PCF25}%
  \BibitemOpen
  \bibfield  {author} {\bibinfo {author} {\bibfnamefont {C.}~\bibnamefont
  {Peters}}, \bibinfo {author} {\bibfnamefont {V.}~\bibnamefont {Cocotos}},\
  and\ \bibinfo {author} {\bibfnamefont {A.}~\bibnamefont {Forbes}},\
  }\bibfield  {title} {\bibinfo {title} {Structured light in atmospheric
  turbulence---a guide to its digital implementation: tutorial},\ }\href
  {https://doi.org/10.1364/AOP.538883} {\bibfield  {journal} {\bibinfo
  {journal} {Adv. Opt. Photon.}\ }\textbf {\bibinfo {volume} {17}},\ \bibinfo
  {pages} {113} (\bibinfo {year} {2025})}\BibitemShut {NoStop}%
\bibitem [{\citenamefont {Martin}\ and\ \citenamefont
  {Flatt\'{e}}(1988)}]{MF88}%
  \BibitemOpen
  \bibfield  {author} {\bibinfo {author} {\bibfnamefont {J.~M.}\ \bibnamefont
  {Martin}}\ and\ \bibinfo {author} {\bibfnamefont {S.~M.}\ \bibnamefont
  {Flatt\'{e}}},\ }\bibfield  {title} {\bibinfo {title} {Intensity images and
  statistics from numerical simulation of wave propagation in 3-d random
  media},\ }\href {https://doi.org/10.1364/AO.27.002111} {\bibfield  {journal}
  {\bibinfo  {journal} {Appl. Opt.}\ }\textbf {\bibinfo {volume} {27}},\
  \bibinfo {pages} {2111} (\bibinfo {year} {1988})}\BibitemShut {NoStop}%
\bibitem [{\citenamefont {Welsh}(1997)}]{B97}%
  \BibitemOpen
  \bibfield  {author} {\bibinfo {author} {\bibfnamefont {B.~M.}\ \bibnamefont
  {Welsh}},\ }\bibfield  {title} {\bibinfo {title} {{Fourier-series-based
  atmospheric phase screen generator for simulating anisoplanatic geometries
  and temporal evolution}},\ }in\ \href {https://doi.org/10.1117/12.279029}
  {\emph {\bibinfo {booktitle} {Propagation and Imaging through the
  Atmosphere}}},\ Vol.\ \bibinfo {volume} {3125},\ \bibinfo {editor} {edited
  by\ \bibinfo {editor} {\bibfnamefont {L.~R.}\ \bibnamefont {Bissonnette}}\
  and\ \bibinfo {editor} {\bibfnamefont {C.}~\bibnamefont {Dainty}}},\ \bibinfo
  {organization} {International Society for Optics and Photonics}\ (\bibinfo
  {publisher} {SPIE},\ \bibinfo {year} {1997})\ pp.\ \bibinfo {pages} {327 --
  338}\BibitemShut {NoStop}%
\bibitem [{\citenamefont {Klen}\ and\ \citenamefont {Semenov}(2023)}]{KS23}%
  \BibitemOpen
  \bibfield  {author} {\bibinfo {author} {\bibfnamefont {M.}~\bibnamefont
  {Klen}}\ and\ \bibinfo {author} {\bibfnamefont {A.~A.}\ \bibnamefont
  {Semenov}},\ }\bibfield  {title} {\bibinfo {title} {Numerical simulations of
  atmospheric quantum channels},\ }\href
  {https://doi.org/10.1103/PhysRevA.108.033718} {\bibfield  {journal} {\bibinfo
   {journal} {Phys. Rev. A}\ }\textbf {\bibinfo {volume} {108}},\ \bibinfo
  {pages} {033718} (\bibinfo {year} {2023})}\BibitemShut {NoStop}%
\bibitem [{\citenamefont {Bachmann}\ \emph {et~al.}(2025)\citenamefont
  {Bachmann}, \citenamefont {Isoard}, \citenamefont {Shatokhin}, \citenamefont
  {Sorelli},\ and\ \citenamefont {Buchleitner}}]{BIS25}%
  \BibitemOpen
  \bibfield  {author} {\bibinfo {author} {\bibfnamefont {D.}~\bibnamefont
  {Bachmann}}, \bibinfo {author} {\bibfnamefont {M.}~\bibnamefont {Isoard}},
  \bibinfo {author} {\bibfnamefont {V.}~\bibnamefont {Shatokhin}}, \bibinfo
  {author} {\bibfnamefont {G.}~\bibnamefont {Sorelli}},\ and\ \bibinfo {author}
  {\bibfnamefont {A.}~\bibnamefont {Buchleitner}},\ }\bibfield  {title}
  {\bibinfo {title} {Accurate zernike-corrected phase screens for arbitrary
  power spectra},\ }\href {https://doi.org/10.1117/1.OE.64.5.058102} {\bibfield
   {journal} {\bibinfo  {journal} {Opt. Engineering}\ }\textbf {\bibinfo
  {volume} {64}},\ \bibinfo {pages} {058102} (\bibinfo {year}
  {2025})}\BibitemShut {NoStop}%
\bibitem [{\citenamefont {Zhou}\ \emph {et~al.}(2015)\citenamefont {Zhou},
  \citenamefont {Zong},\ and\ \citenamefont {Liu}}]{ZZL15}%
  \BibitemOpen
  \bibfield  {author} {\bibinfo {author} {\bibfnamefont {J.}~\bibnamefont
  {Zhou}}, \bibinfo {author} {\bibfnamefont {J.}~\bibnamefont {Zong}},\ and\
  \bibinfo {author} {\bibfnamefont {D.}~\bibnamefont {Liu}},\ }\bibfield
  {title} {\bibinfo {title} {Coupled mode theory for orbital angular momentum
  modes transmission in the presence of atmosphere turbulence},\ }\href
  {https://doi.org/10.1364/OE.23.031964} {\bibfield  {journal} {\bibinfo
  {journal} {Opt. Express}\ }\textbf {\bibinfo {volume} {23}},\ \bibinfo
  {pages} {31964} (\bibinfo {year} {2015})}\BibitemShut {NoStop}%
\bibitem [{\citenamefont {Zhou}\ \emph {et~al.}(2016)\citenamefont {Zhou},
  \citenamefont {Zong},\ and\ \citenamefont {Liu}}]{ZZL16}%
  \BibitemOpen
  \bibfield  {author} {\bibinfo {author} {\bibfnamefont {J.}~\bibnamefont
  {Zhou}}, \bibinfo {author} {\bibfnamefont {J.}~\bibnamefont {Zong}},\ and\
  \bibinfo {author} {\bibfnamefont {D.}~\bibnamefont {Liu}},\ }\bibfield
  {title} {\bibinfo {title} {The higher order statistics of {OAM} modal
  amplitudes under atmosphere turbulence},\ }\href
  {https://doi.org/10.1109/LPT.2016.2615036} {\bibfield  {journal} {\bibinfo
  {journal} {IEEE Photon. Technol. Lett.}\ }\textbf {\bibinfo {volume} {28}},\
  \bibinfo {pages} {2653} (\bibinfo {year} {2016})}\BibitemShut {NoStop}%
\bibitem [{\citenamefont {Zhou}\ \emph {et~al.}(2018)\citenamefont {Zhou},
  \citenamefont {Wu}, \citenamefont {Zong},\ and\ \citenamefont {Hu}}]{ZWZ18}%
  \BibitemOpen
  \bibfield  {author} {\bibinfo {author} {\bibfnamefont {J.}~\bibnamefont
  {Zhou}}, \bibinfo {author} {\bibfnamefont {J.}~\bibnamefont {Wu}}, \bibinfo
  {author} {\bibfnamefont {J.}~\bibnamefont {Zong}},\ and\ \bibinfo {author}
  {\bibfnamefont {Q.}~\bibnamefont {Hu}},\ }\bibfield  {title} {\bibinfo
  {title} {Optimal mode set selection for free space optical communications in
  the presence of atmosphere turbulence},\ }\href
  {https://doi.org/10.1109/JLT.2018.2808912} {\bibfield  {journal} {\bibinfo
  {journal} {J. Lightwav. Technol.}\ }\textbf {\bibinfo {volume} {36}},\
  \bibinfo {pages} {2222} (\bibinfo {year} {2018})}\BibitemShut {NoStop}%
\bibitem [{\citenamefont {Roux}\ \emph {et~al.}(2015)\citenamefont {Roux},
  \citenamefont {Wellens},\ and\ \citenamefont {Shatokhin}}]{RWS15}%
  \BibitemOpen
  \bibfield  {author} {\bibinfo {author} {\bibfnamefont {F.~S.}\ \bibnamefont
  {Roux}}, \bibinfo {author} {\bibfnamefont {T.}~\bibnamefont {Wellens}},\ and\
  \bibinfo {author} {\bibfnamefont {V.~N.}\ \bibnamefont {Shatokhin}},\
  }\bibfield  {title} {\bibinfo {title} {Entanglement evolution of twisted
  photons in strong atmospheric turbulence},\ }\href
  {https://doi.org/10.1103/PhysRevA.92.012326} {\bibfield  {journal} {\bibinfo
  {journal} {Phys. Rev. A}\ }\textbf {\bibinfo {volume} {92}},\ \bibinfo
  {pages} {012326} (\bibinfo {year} {2015})}\BibitemShut {NoStop}%
\bibitem [{\citenamefont {Bachmann}\ \emph {et~al.}(2019)\citenamefont
  {Bachmann}, \citenamefont {Shatokhin},\ and\ \citenamefont
  {Buchleitner}}]{BSB19}%
  \BibitemOpen
  \bibfield  {author} {\bibinfo {author} {\bibfnamefont {D.}~\bibnamefont
  {Bachmann}}, \bibinfo {author} {\bibfnamefont {V.~N.}\ \bibnamefont
  {Shatokhin}},\ and\ \bibinfo {author} {\bibfnamefont {A.}~\bibnamefont
  {Buchleitner}},\ }\bibfield  {title} {\bibinfo {title} {Universal
  entanglement decay of photonic orbital angular momentum qubit states in
  atmospheric turbulence: an analytical treatment},\ }\href
  {https://doi.org/10.1088/1751-8121/ab3f3c} {\bibfield  {journal} {\bibinfo
  {journal} {Journal of Physics A: Mathematical and Theoretical}\ }\textbf
  {\bibinfo {volume} {52}},\ \bibinfo {pages} {405303} (\bibinfo {year}
  {2019})}\BibitemShut {NoStop}%
\bibitem [{\citenamefont {Kravtsov}\ \emph {et~al.}(2018)\citenamefont
  {Kravtsov}, \citenamefont {Zhutov}, \citenamefont {Radchenko},\ and\
  \citenamefont {Kulik}}]{KZR18}%
  \BibitemOpen
  \bibfield  {author} {\bibinfo {author} {\bibfnamefont {K.~S.}\ \bibnamefont
  {Kravtsov}}, \bibinfo {author} {\bibfnamefont {A.~K.}\ \bibnamefont
  {Zhutov}}, \bibinfo {author} {\bibfnamefont {I.~V.}\ \bibnamefont
  {Radchenko}},\ and\ \bibinfo {author} {\bibfnamefont {S.~P.}\ \bibnamefont
  {Kulik}},\ }\bibfield  {title} {\bibinfo {title} {Turbulence-induced optical
  loss and cross-talk in spatial-mode multiplexed or single-mode free-space
  communication channels},\ }\href {https://doi.org/10.1103/PhysRevA.98.063831}
  {\bibfield  {journal} {\bibinfo  {journal} {Phys. Rev. A}\ }\textbf {\bibinfo
  {volume} {98}},\ \bibinfo {pages} {063831} (\bibinfo {year}
  {2018})}\BibitemShut {NoStop}%
\bibitem [{\citenamefont {Cox}\ \emph {et~al.}(2016)\citenamefont {Cox},
  \citenamefont {Rosales-Guzm\'{a}n}, \citenamefont {Lavery}, \citenamefont
  {Versfeld},\ and\ \citenamefont {Forbes}}]{CRL16}%
  \BibitemOpen
  \bibfield  {author} {\bibinfo {author} {\bibfnamefont {M.~A.}\ \bibnamefont
  {Cox}}, \bibinfo {author} {\bibfnamefont {C.}~\bibnamefont
  {Rosales-Guzm\'{a}n}}, \bibinfo {author} {\bibfnamefont {M.~P.~J.}\
  \bibnamefont {Lavery}}, \bibinfo {author} {\bibfnamefont {D.~J.}\
  \bibnamefont {Versfeld}},\ and\ \bibinfo {author} {\bibfnamefont
  {A.}~\bibnamefont {Forbes}},\ }\bibfield  {title} {\bibinfo {title} {On the
  resilience of scalar and vector vortex modes in turbulence},\ }\href
  {https://doi.org/10.1364/OE.24.018105} {\bibfield  {journal} {\bibinfo
  {journal} {Opt. Express}\ }\textbf {\bibinfo {volume} {24}},\ \bibinfo
  {pages} {18105} (\bibinfo {year} {2016})}\BibitemShut {NoStop}%
\bibitem [{\citenamefont {Mphuthi}\ \emph {et~al.}(2018)\citenamefont
  {Mphuthi}, \citenamefont {Botha},\ and\ \citenamefont {Forbes}}]{MBF18}%
  \BibitemOpen
  \bibfield  {author} {\bibinfo {author} {\bibfnamefont {N.}~\bibnamefont
  {Mphuthi}}, \bibinfo {author} {\bibfnamefont {R.}~\bibnamefont {Botha}},\
  and\ \bibinfo {author} {\bibfnamefont {A.}~\bibnamefont {Forbes}},\
  }\bibfield  {title} {\bibinfo {title} {Are {B}essel beams resilient to
  aberrations and turbulence?},\ }\href
  {https://doi.org/10.1364/JOSAA.35.001021} {\bibfield  {journal} {\bibinfo
  {journal} {J. Opt. Soc. Am. A}\ }\textbf {\bibinfo {volume} {35}},\ \bibinfo
  {pages} {1021} (\bibinfo {year} {2018})}\BibitemShut {NoStop}%
\bibitem [{\citenamefont {Beland}(1993)}]{B93}%
  \BibitemOpen
  \bibfield  {author} {\bibinfo {author} {\bibfnamefont {R.~R.}\ \bibnamefont
  {Beland}},\ }\bibinfo {title} {Propagation through atmospheric optical
  turbulence},\ in\ \href {https://doi.org/10.1117/3.2543821.ch2} {\emph
  {\bibinfo {booktitle} {The Infrared \& Electro-Optical Systems Handbook,
  Volume 2}}},\ \bibinfo {editor} {edited by\ \bibinfo {editor} {\bibfnamefont
  {F.~G.}\ \bibnamefont {Smith}}}\ (\bibinfo  {publisher} {SPIE Press / The
  Infrared Information Analysis Center},\ \bibinfo {year} {1993})\ pp.\
  \bibinfo {pages} {157--232},\ \bibinfo {note} {chapter 2}\BibitemShut
  {NoStop}%
\bibitem [{\citenamefont {Lawrence}\ and\ \citenamefont
  {Strohbehn}(1970)}]{LS70}%
  \BibitemOpen
  \bibfield  {author} {\bibinfo {author} {\bibfnamefont {R.}~\bibnamefont
  {Lawrence}}\ and\ \bibinfo {author} {\bibfnamefont {J.}~\bibnamefont
  {Strohbehn}},\ }\bibfield  {title} {\bibinfo {title} {A survey of clear-air
  propagation effects relevant to optical communications},\ }\href
  {https://doi.org/10.1109/PROC.1970.7977} {\bibfield  {journal} {\bibinfo
  {journal} {Proceedings of the IEEE}\ }\textbf {\bibinfo {volume} {58}},\
  \bibinfo {pages} {1523} (\bibinfo {year} {1970})}\BibitemShut {NoStop}%
\bibitem [{\citenamefont {Fried}(1966)}]{F66}%
  \BibitemOpen
  \bibfield  {author} {\bibinfo {author} {\bibfnamefont {D.~L.}\ \bibnamefont
  {Fried}},\ }\bibfield  {title} {\bibinfo {title} {Optical resolution through
  a randomly inhomogeneous medium for very long and very short exposures},\
  }\href {https://doi.org/10.1364/JOSA.56.001372} {\bibfield  {journal}
  {\bibinfo  {journal} {J. Opt. Soc. Am.}\ }\textbf {\bibinfo {volume} {56}},\
  \bibinfo {pages} {1372} (\bibinfo {year} {1966})}\BibitemShut {NoStop}%
\bibitem [{\citenamefont {Roddier}(1981)}]{R81}%
  \BibitemOpen
  \bibfield  {author} {\bibinfo {author} {\bibfnamefont {F.}~\bibnamefont
  {Roddier}},\ }\bibfield  {title} {\bibinfo {title} {V the effects of
  atmospheric turbulence in optical astronomy}\ }(\bibinfo  {publisher}
  {Elsevier},\ \bibinfo {year} {1981})\ pp.\ \bibinfo {pages}
  {281--376}\BibitemShut {NoStop}%
\bibitem [{\citenamefont {Tatarski}(1971)}]{T71}%
  \BibitemOpen
  \bibfield  {author} {\bibinfo {author} {\bibfnamefont {V.~I.}\ \bibnamefont
  {Tatarski}},\ }\href@noop {} {\emph {\bibinfo {title} {The Effects of the
  Turbulent Atmosphere on Wave Propagation}}}\ (\bibinfo  {publisher} {Israel
  Program for Scientific Translations.},\ \bibinfo {address} {Jerusalem,
  Israel},\ \bibinfo {year} {1971})\BibitemShut {NoStop}%
\bibitem [{\citenamefont {Danakas}\ and\ \citenamefont {Aravind}(1992)}]{DA92}%
  \BibitemOpen
  \bibfield  {author} {\bibinfo {author} {\bibfnamefont {S.}~\bibnamefont
  {Danakas}}\ and\ \bibinfo {author} {\bibfnamefont {P.~K.}\ \bibnamefont
  {Aravind}},\ }\bibfield  {title} {\bibinfo {title} {Analogies between 2
  optical-systems (photon-beam splitters and laser-beams) and 2 quantum-systems
  (the 2-dimensional oscillator and the 2-dimensional hydrogen-atom)},\ }\href
  {https://doi.org/10.1103/PhysRevA.45.1973} {\bibfield  {journal} {\bibinfo
  {journal} {Phys. Rev. A}\ }\textbf {\bibinfo {volume} {45}},\ \bibinfo
  {pages} {1973} (\bibinfo {year} {1992})}\BibitemShut {NoStop}%
\bibitem [{\citenamefont {Restuccia}\ \emph {et~al.}(2016)\citenamefont
  {Restuccia}, \citenamefont {Giovannini}, \citenamefont {Gibson},\ and\
  \citenamefont {Padgett}}]{RGG16}%
  \BibitemOpen
  \bibfield  {author} {\bibinfo {author} {\bibfnamefont {S.}~\bibnamefont
  {Restuccia}}, \bibinfo {author} {\bibfnamefont {D.}~\bibnamefont
  {Giovannini}}, \bibinfo {author} {\bibfnamefont {G.}~\bibnamefont {Gibson}},\
  and\ \bibinfo {author} {\bibfnamefont {M.}~\bibnamefont {Padgett}},\
  }\bibfield  {title} {\bibinfo {title} {Comparing the information capacity of
  {L}aguerre-{G}aussian and {H}ermite-{G}aussian modal sets in a
  finite-aperture system},\ }\href {https://doi.org/10.1364/OE.24.027127}
  {\bibfield  {journal} {\bibinfo  {journal} {Opt. Express}\ }\textbf {\bibinfo
  {volume} {24}},\ \bibinfo {pages} {27127} (\bibinfo {year}
  {2016})}\BibitemShut {NoStop}%
\bibitem [{\citenamefont {Gradshteyn}\ and\ \citenamefont
  {Ryzhik}(2007)}]{GRXX}%
  \BibitemOpen
  \bibfield  {author} {\bibinfo {author} {\bibfnamefont {I.~S.}\ \bibnamefont
  {Gradshteyn}}\ and\ \bibinfo {author} {\bibfnamefont {I.~M.}\ \bibnamefont
  {Ryzhik}},\ }\href@noop {} {\emph {\bibinfo {title} {Table of Integrals,
  Series, and Products}}},\ \bibinfo {edition} {7th}\ ed.,\ edited by\ \bibinfo
  {editor} {\bibfnamefont {A.}~\bibnamefont {Jeffrey}}\ and\ \bibinfo {editor}
  {\bibfnamefont {D.}~\bibnamefont {Zwillinger}}\ (\bibinfo  {publisher}
  {Academic Press},\ \bibinfo {address} {Amsterdam},\ \bibinfo {year}
  {2007})\BibitemShut {NoStop}%
\bibitem [{\citenamefont {Jarzynski}(1997)}]{J97}%
  \BibitemOpen
  \bibfield  {author} {\bibinfo {author} {\bibfnamefont {C.}~\bibnamefont
  {Jarzynski}},\ }\bibfield  {title} {\bibinfo {title} {Equilibrium free-energy
  differences from nonequilibrium measurements: A master-equation approach},\
  }\href {https://doi.org/10.1103/PhysRevE.56.5018} {\bibfield  {journal}
  {\bibinfo  {journal} {Phys. Rev. E}\ }\textbf {\bibinfo {volume} {56}},\
  \bibinfo {pages} {5018} (\bibinfo {year} {1997})}\BibitemShut {NoStop}%
\bibitem [{\citenamefont {Sohl-Dickstein}\ \emph {et~al.}(2015)\citenamefont
  {Sohl-Dickstein}, \citenamefont {Weiss}, \citenamefont {Maheswaranathan},\
  and\ \citenamefont {Ganguli}}]{SWM15}%
  \BibitemOpen
  \bibfield  {author} {\bibinfo {author} {\bibfnamefont {J.}~\bibnamefont
  {Sohl-Dickstein}}, \bibinfo {author} {\bibfnamefont {E.}~\bibnamefont
  {Weiss}}, \bibinfo {author} {\bibfnamefont {N.}~\bibnamefont
  {Maheswaranathan}},\ and\ \bibinfo {author} {\bibfnamefont {S.}~\bibnamefont
  {Ganguli}},\ }\bibfield  {title} {\bibinfo {title} {Deep unsupervised
  learning using nonequilibrium thermodynamics},\ }in\ \href@noop {} {\emph
  {\bibinfo {booktitle} {International Conference on Machine Learning}}}\
  (\bibinfo {address} {Lille, France},\ \bibinfo {year} {2015})\ pp.\ \bibinfo
  {pages} {2256–--2265}\BibitemShut {NoStop}%
\bibitem [{\citenamefont {Owen}\ and\ \citenamefont {Horowitz}(2023)}]{OH23}%
  \BibitemOpen
  \bibfield  {author} {\bibinfo {author} {\bibfnamefont {J.~A.}\ \bibnamefont
  {Owen}}\ and\ \bibinfo {author} {\bibfnamefont {J.~M.}\ \bibnamefont
  {Horowitz}},\ }\bibfield  {title} {\bibinfo {title} {Size limits the
  sensitivity of kinetic schemes},\ }\href
  {https://doi.org/10.1038/s41467-023-36705-8} {\bibfield  {journal} {\bibinfo
  {journal} {Nature Commun.}\ }\textbf {\bibinfo {volume} {14}},\ \bibinfo
  {pages} {1280} (\bibinfo {year} {2023})}\BibitemShut {NoStop}%
\end{thebibliography}
\end{document}